\documentclass[reprint,floatfix]{revtex4-2}

\usepackage{url}
\usepackage{amsmath}
\usepackage{graphicx}
\usepackage{dcolumn}
\usepackage{lineno}
\usepackage{physics}
\usepackage{bm}
\usepackage{hyperref}
\usepackage{epsfig,colordvi}
\usepackage{hyperref}
\usepackage{graphicx}
\usepackage{dcolumn}
\usepackage{xspace}
\usepackage{times}
\usepackage{bm}
\usepackage{soul}
\usepackage{ulem}
\usepackage{xcolor}
\usepackage{color}
\usepackage{natbib}
\usepackage{float}
\usepackage{booktabs} 

 \def\be{\begin{equation}}
 \def\ee{\end{equation}}
 \def\bea{\begin{eqnarray}}
 \def\eea{\end{eqnarray}}
 \def\bean{\begin{eqnarray*}}
 \def\eean{\end{eqnarray*}}

\newcommand{\snn}{\mbox{$\sqrt{s_{\rm NN}}$}\xspace}

 \def\be{\begin{equation}}
 \def\ee{\end{equation}}
 \def\bea{\begin{eqnarray}}
 \def\eea{\end{eqnarray}}
 \def\bean{\begin{eqnarray*}}
 \def\eean{\end{eqnarray*}}

\begin{document}

\title{A Data-Guided Coalescence Model for Light Nuclei and Hypernuclei Production in Relativistic Heavy-Ion Collisions at $\sqrt{s_{\rm{NN}}} = 3$--200~GeV}

\author{Yue Hang Leung$^{1}$, Yingjie Zhou$^{2}$, Norbert Herrmann$^{1,2}$}

\affiliation{$^{1}$ University of Heidelberg, Heidelberg 69120, Germany}
\affiliation{$^{2}$ GSI Helmholtzzentrum f\"ur Schwerionenforschung GmbH, Darmstadt 64291, Germany}   


\begin{abstract} 
The production of light hypernuclei in relativistic heavy-ion collisions provides a unique opportunity to probe hyperon--nucleon interactions and possible three-body forces, which are central to the resolution of the hyperon puzzle in neutron star matter. In this work, we develop a data-guided coalescence framework in which the source size is extracted from proton and deuteron yields and used, together with measured proton and $\Lambda$ spectra, to predict the production of $A=3$ nuclei $(t,{}^{3}\rm{He})$ and hypernuclei $({}^{3}_{\Lambda}\rm{H})$ in $\sqrt{s_{\rm{NN}}}=3$--200~GeV Au+Au collisions. For tritons, calculations with a Gaussian wave function reproduce experimental spectra and yield ratios across this energy range. For the hypertriton, the model predictions are highly sensitive to the assumed wave function. This sensitivity is strongest at low collision energies and in low-multiplicity environments, implying that such conditions are particularly valuable for probing hypernuclear structure. 
\end{abstract}

\maketitle

\section{Introduction}
\label{sec:introduction}

The hyperon puzzle is a central problem in modern nuclear astrophysics, arising from the theoretical expectation that hyperons---baryons containing strange quarks---should appear in the dense cores of neutron stars once the baryon chemical potential becomes sufficiently high. Their presence softens the equation of state (EoS) of dense matter, leading to a significant reduction in the predicted maximum neutron star mass~\cite{Prakash:1996xs}. This theoretical prediction, however, conflicts with observations of massive neutron stars with masses around $2\,M_\odot$~\cite{Demorest:2010bx, Antoniadis:2013pzd, NANOGrav:2019jur}, which demand a sufficiently stiff EoS to support such stars against gravitational collapse. Reconciling the existence of hyperons with these astrophysical observations constitutes the essence of the hyperon puzzle.

A crucial element in addressing this problem lies in the treatment of hyperon--nucleon (YN) interactions. These interactions, which describe the forces between hyperons and nucleons, are relatively well constrained at low energies by experimental data from hypernuclei. They typically feature an attractive component, which facilitates the onset of hyperons at densities only a few times higher than nuclear saturation density. However, the extrapolation of YN interactions to the extreme densities found in neutron star cores is highly uncertain, and their detailed structure strongly influences the density threshold and abundance of hyperons.

To further refine the description of dense matter, many studies have emphasized the importance of three-body forces. In nuclear physics, three-nucleon forces are known to play an essential role in reproducing the saturation properties of nuclear matter and the structure of light nuclei~\cite{Epelbaum:2008ga, Barrett:2013nh, Hagen:2012fb, Hagen:2013nca, Soma:2012zd, Roth:2011ar, Stroberg:2019mxo,Carlson:2014vla}. Extending this concept to systems involving strangeness, one must also consider hyperon--nucleon--nucleon (YNN) and potentially hyperon--hyperon--nucleon (YYN) interactions. These three-body forces may be repulsive at high densities~\cite{Gerstung:2020ktv, Lonardoni:2014bwa}, thereby providing additional pressure support in the stellar core. As a result, they can counteract the softening caused by hyperons, thus offering a promising route toward resolving the hyperon puzzle. The quantitative determination of these forces, however, remains challenging due to the scarcity of experimental data and the theoretical complexity of many-body calculations in the strangeness sector.

Hypernuclei, which contain at least one hyperon bound to nucleons, are abundantly produced in relativistic heavy-ion collisions and provide valuable information on YN and three-body YNN interactions~\cite{Hildenbrand:2019sgp, Le:2024aox, Haidenbauer:2025zrr, Kohno:2022bsq, Kamada:2023txx}. Traditionally, constraints on three-body hyperonic forces have primarily been obtained from hypernuclear binding energies, which probe the underlying interactions in the low-density nuclear regime~\cite{Tong:2025fzv, Le:2024rkd}. While the hyperon puzzle manifests itself at densities well above nuclear saturation in neutron-star interiors, its resolution relies on accurate knowledge of the underlying YN and YNN interactions across all density regimes, for which hypernuclear observables serve as essential low-density benchmarks. In parallel, theoretical efforts have aimed at developing $\textit{ab initio}$ frameworks applicable to both hypernuclei and dense matter relevant for neutron stars~\cite{Tong:2025fzv, Haidenbauer:2025zrr}, within which the same microscopic interactions may also determine hypernuclear wave functions. In recent years, increasingly precise measurements of hypernuclear production yields have become available, opening new opportunities to investigate these interactions through the dynamical formation of hypernuclei. In this work, we explore whether hypernuclear yield measurements from relativistic heavy-ion collisions can provide complementary constraints on such interactions by studying the sensitivity of production yields to the wave function of the hypertriton $(^{3}_{\Lambda}\mathrm{H})$—the lightest known hypernucleus consisting of a proton, neutron, and $\Lambda$ hyperon.

There are two main theoretical approaches to calculate nuclei and hypernuclei yields in relativistic heavy-ion collisions: the thermal model~\cite{Andronic:2018qqt,Cleymans:2005xv,Andronic:2010qu} and the coalescence model~\cite{Scheibl:1998tk, Steinheimer:2012tb, Aichelin:1991xy, Bleicher:1995dw, Nagle:1996vp, Kittiratpattana:2020daw, Botvina:2014lga}. While the thermal model assumes chemical equilibrium at freeze-out, the coalescence picture builds the yield from the overlap of nucleon and hyperon sources with the internal wave function of the composite nucleus. Increasing evidence~\cite{STAR:2022hbp,STAR:2022fnj, ALICE:2022veq,ALICE:2025byl} suggests that the coalescence approach provides a more accurate description of nuclei and hypernuclei production.

Within the coalescence framework, the yield---which is directly related to the coalescence probability---is essentially a convolution of the hypernuclear wave function with the phase-space distribution of the nucleons and hyperons at freeze-out~\cite{Sato:1981ez}. This implies that precise yield measurements can, in principle, encode direct information about the internal wave function of hypernuclei. For the $^{3}_{\Lambda}\rm{H}$, this is especially significant, since its wave function remains poorly constrained. Consequently, one of the main goals of this work is to explore the possibility of using yield measurements to extract information about the $^{3}_{\Lambda}\rm{H}$ wave function, thereby providing new constraints on YN interactions and possibly three-body forces.

The STAR Collaboration has collected data from the Beam Energy Scan II (BES-II) program in 2019-2021, which covers the energy range $3$--$27$~GeV. At \snn=$200$~GeV, large-statistics data sets are also available from Zr+Zr and Ru+Ru collisions taken in 2018. Furthermore, future facilities such as the China HyperNuclear Spectrometer (CHNS) at the High Intensity heavy-ion Accelerator Facility~\cite{Zhou:2022pxl}  and CBM detector at FAIR~\cite{CBM:2016kpk}  will provide U+U and Au+Au collisions at around \snn=3--$5$ GeV. For these reasons, our analysis will focus on the energy range of $\sqrt{s_{\rm{NN}}} = 3$--$200$~GeV, where current and upcoming experimental opportunities are particularly rich. 

The remainder of this paper is organized as follows. In Sec.~\ref{sec:methods} we describe the methodology of the coalescence calculations employed in this study. Secs.~\ref{sec:wavefunction} and~\ref{sec:source} introduce the inputs to the data-driven coalescence approach, namely the wave functions of light nuclei and hypernuclei, and the parameterization of the nucleon and hyperon source distributions, respectively. Our main results are presented in Sec.~\ref{sec:results}, where we focus on triton, $^3$He, and $^{3}_{\Lambda}\rm{H}$ production, including transverse-momentum ($p_T$) spectra, $p_T$-integrated yield ratios, and mean transverse momenta. Finally, we summarize our findings and provide an outlook in Sec.~\ref{sec:summary}. 

\section{Methods}
\label{sec:methods}

In this work, we adopt a data-driven approach to calculate the production of light nuclei and hypernuclei within the coalescence framework. A commonly used strategy in the field is to combine microscopic transport models, which simulate the dynamical evolution of the collision, with a coalescence “afterburner” that estimates cluster formation at kinetic freeze-out. While this transport-based framework has been widely applied~\cite{Zhao:2021dka, Reichert:2022mek, Steinheimer:2012tb, Liu:2024ilw}, it suffers from an important limitation: the underlying transport model may not accurately reproduce the absolute yields or the detailed phase-space distributions of nucleons. Since the coalescence probability is highly sensitive to the shape and normalization of these distributions, such deficiencies can significantly bias the predicted hypernuclear yields.

This issue becomes particularly severe at low collision energies, where both particle yields and phase-space distributions of nucleons at freeze-out are strongly influenced by the EoS of dense nuclear matter~\cite{Zhou:2025zgn, Kireyeu:2024hjo}. The EoS in this regime remains poorly constrained, leading to large theoretical uncertainties when transport models are used as input. To mitigate these issues, we instead pursue a data-driven approach, making maximal use of available experimental data to constrain the nucleon and hyperon distributions that enter the coalescence formalism. In this way, the sensitivity of our results to model-dependent assumptions about the EoS is reduced, and a more direct connection between measured observables and the extracted hypernuclear properties may be established.

We first introduce the coalescence parameters $B_A$ for a (hyper)nuclei $X$ with mass number $A$, defined as
\begin{equation}
B_A(X) = \frac{E_{A}\frac{d^3N_A}{d^3P}}
{\prod_{i=1}^A \left( E_{N_{i}}\frac{dN_{N_i}}{d^3p_i} \right)} \,,
\label{eq:bn}
\end{equation}
where $N_{i}$ are the constituent nucleons or hyperons, and $E_{A}\frac{d^3N_A}{d^3P}$ and $E_{N_{i}}\frac{dN_{N_i}}{d^3p_i}$ are the invariant yields of the (hyper)nuclei and nucleon/hyperon respectively. The coalescence parameters encapsulate the probability of coalescence formation, and allows one to compute the nuclei yields so long as the constitent nucleon yields are known. 

Our methodology follows the \emph{coalescence--correlation framework} of Ref.~\cite{Bellini:2020cbj}, which unifies the description of femtoscopic correlations between nucleons and the coalescence model for light (hyper)nuclei production. Below is a brief review of the key concepts, for detailed derivations please refer to Ref.~\cite{Bellini:2020cbj}

The description of femtoscopic correlations between nucleons \cite{Koonin:1977fh} and the coalescence model for nuclei \cite{Bond:1977fd, Sato:1981ez} are two aspects of the same theoretical framework. The coalescence--correlation link can be derived from a relativistic Bethe--Salpeter formalism~\cite{Schweber:1955zz}, reducing to an effective quantum-mechanical picture in the source rest frame. A key assumption is the factorization of the density matrix into single-particle components. Under the smoothness approximation and the equal-time approximation, $B_A$ can be expressed as an overlap between the multi-particle 
source function and the nuclear wave function. Explicitly,

\begin{align}
B_2(d) \approx& \frac{2(2s_d + 1)}{m\,(2s_N + 1)^2 (2\pi)^3} 
\int d^3r \; \lvert \Phi_d(r)\rvert^2\, \mathcal{S}_2(r), \label{eq:b2}\\
B_3({}^{3}{\rm{He}}) \approx& \frac{3}{m^2} \frac{2s_{\mathrm{^3He}} + 1}{(2s_N + 1)^3} 
(2\pi)^6
\int d^3r_{pp} \int d^3r_n \nonumber\\
&\lvert \Phi_{^3\mathrm{He}}(r_{pp}, r_n)\rvert^2 \, 
\mathcal{S}_3(r_{pp}, r_n), \label{eq:b3} \\
B_{3}(^{3}_{\Lambda}{\rm{H}}) \approx& \frac{3}{m^2} \frac{2s_{^{3}_{\Lambda}{\rm{H}}} + 1}{(2s_N + 1)^3} 
(2\pi)^6\int d^3r_{pn} \int d^3r_{\Lambda} \nonumber \\
&\lvert \Phi_{^3_{\Lambda}{\rm{H}}}(r_{pn}, r_{\Lambda})\rvert^2 \, \mathcal{S}_{3}(r_{pn}, r_{\Lambda}), \label{eq:b3lambda}
\end{align}

\noindent where $\Phi_d$, $\Phi_{^3\mathrm{He}}$, and $\Phi_{^3_{\Lambda}{\rm{H}}}$ are the 
corresponding nuclear wave functions, $s_{i}$ is the spin of the particle $i$, and $\mathcal{S}_2$ and $\mathcal{S}_{3}$ are the two-particle and three-particle source functions respectively.

Assuming an isotropic Gaussian emission source with radius parameter $R_{\mathrm{inv}}$, 
the two- and three-particle source functions take the forms
\begin{align}
\mathcal{S}_{2}(r) &= \frac{1}{(4\pi R^2_{\mathrm{inv}})^{3/2}}
\exp\!\left(-\frac{r^2}{4R_{\mathrm{inv}}^2}\right), \\
\mathcal{S}_{3}(r_{12}, r_3) &= \frac{1}{(12\pi^{2} R^4_{\mathrm{inv}})^{3/2}}
\exp\!\left(-\frac{r_{12}^2+\tfrac{4}{3}r_3^2}{4R_{\mathrm{inv}}^2}\right),
\end{align}
where $r_{12}$ is the pair separation and $r_3$ is the distance of the third 
particle from the pair center-of-mass.

With the above expressions, the evaluation of the coalescence parameters can be achieved so long as the nuclear wave functions and $\Phi_{{}^3_\Lambda \rm{H}}$, and the source radius $R_{\rm inv}$ are provided. The determination of these inputs will be discussed in the following section.

\subsection{Inputs from Thermal Model}

\begin{figure}[h!] 
    \centering
    \includegraphics[width=.49\linewidth]{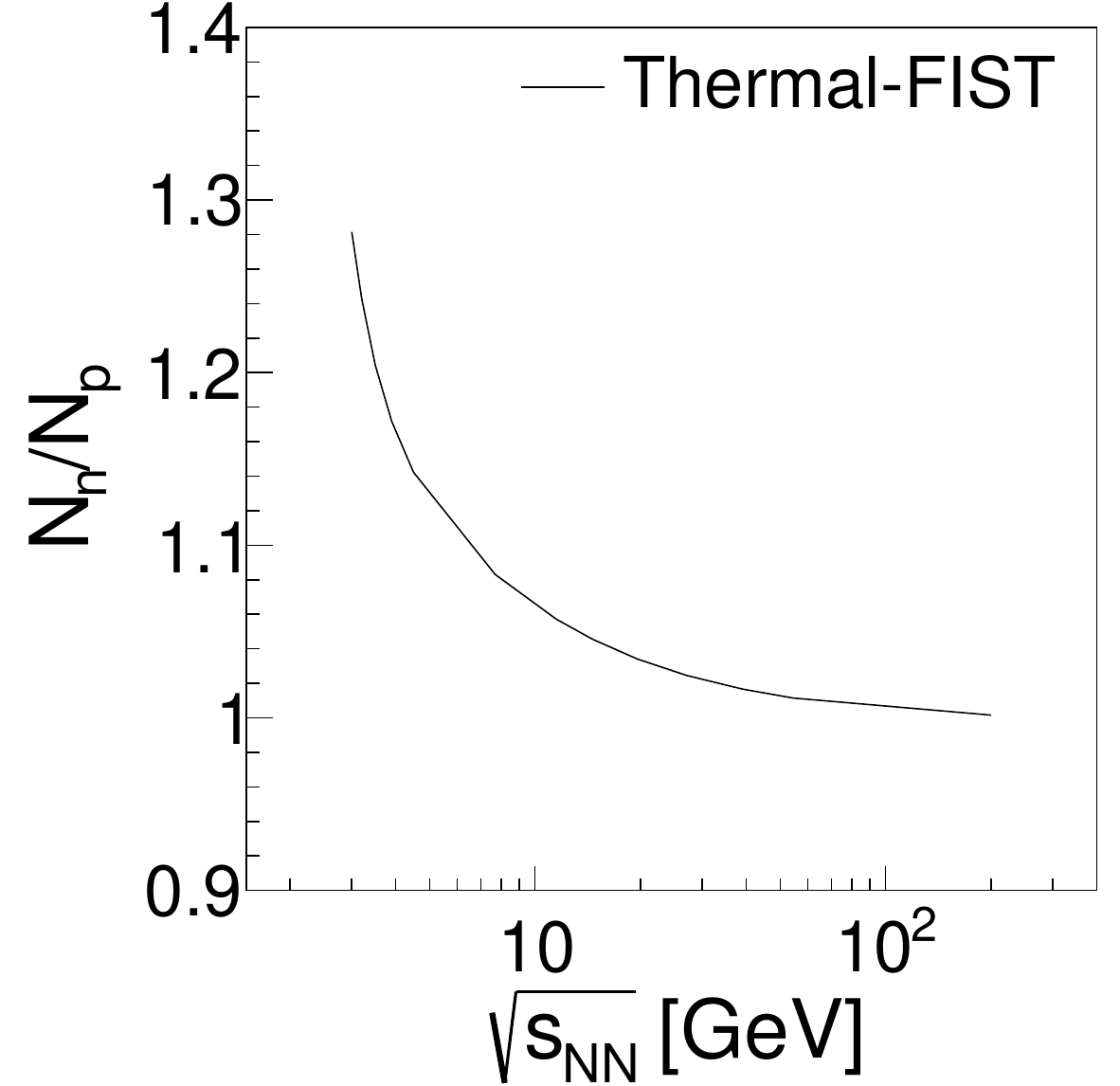} 
    \includegraphics[width=.49\linewidth]{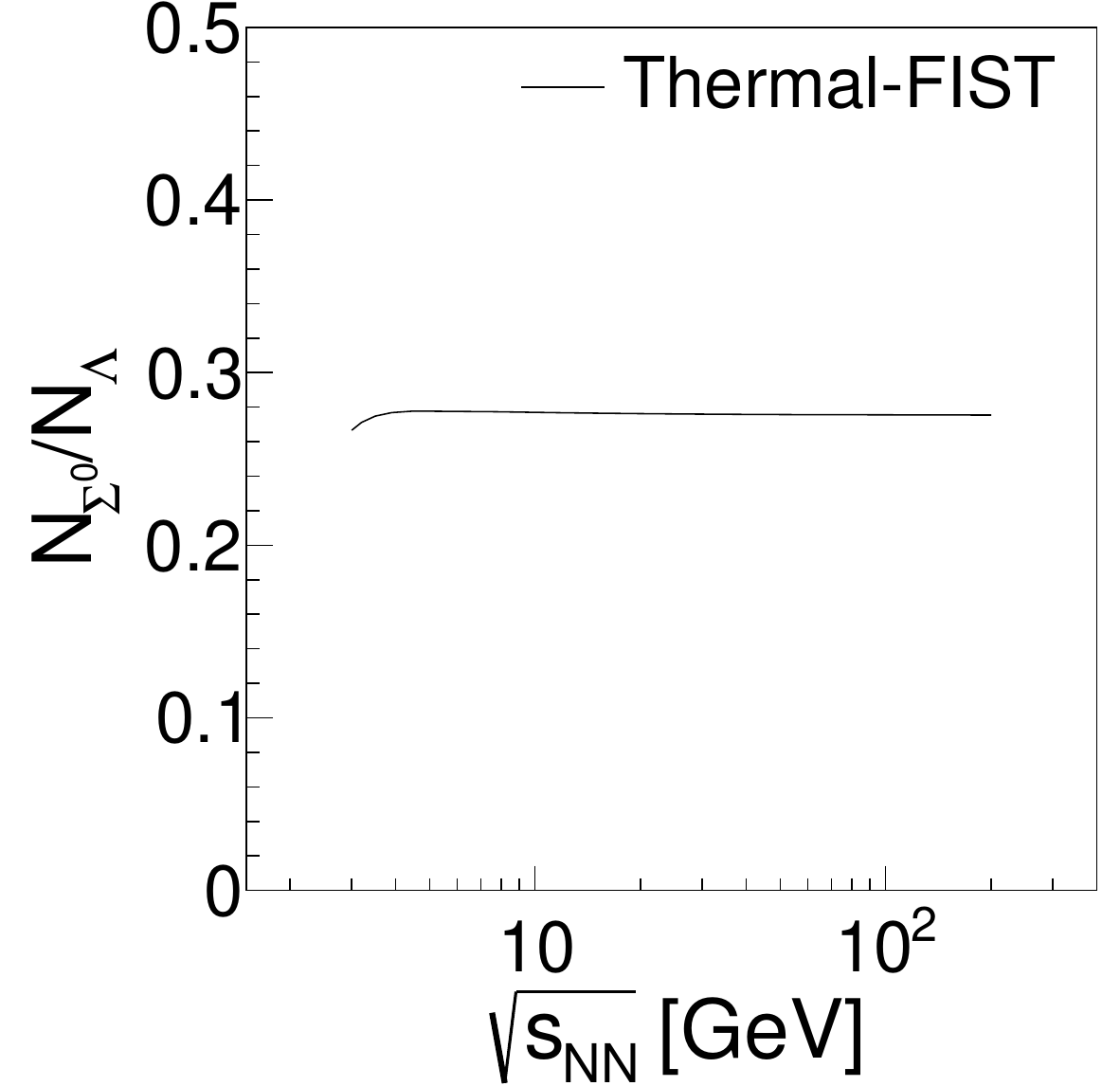} 
    \caption{Yield ratios $N_{n}/N_{p}$ (left) and $N_{\Sigma^{0}}/N_{\Lambda}$ (right) as a function of collision energy as predicted by Thermal-FIST~\cite{Vovchenko:2019pjl}.}
    \label{fig:n_over_p}
\end{figure}

Our main objective is to apply the coalescence methodology to Au+Au collisions over the energy range $\sqrt{s_{\rm{NN}}}=3$--200~GeV. To this end, we implement two key modifications to the original coalescence calculation:

\textbf{(i) Neutron spectra estimation.}  
In the original approach, the neutron spectra were assumed to be identical to the proton spectra. However, in low-energy Au+Au collisions the neutron yield exceeds the proton yield due to the neutron-rich composition of gold nuclei and baryon stopping. To account for this, we abandon the equality assumption and instead estimate the neutron yield from the measured proton yield combined with the neutron-to-proton ratio obtained from thermal model calculations. The \textsc{Thermal-FIST} package~\cite{Vovchenko:2019pjl} is employed, with freeze-out parameters $(T,\mu_B)$ taken from Ref.~\cite{Vovchenko:2015idt}. Further details are provided in Appendix~\ref{app:thermal}. The neutron-to-proton ratio $N_{n}/N_{p}$ as a function of collision energy in Au+Au collisions is shown in Fig.~\ref{fig:n_over_p}.

\textbf{(ii) Treatment of chaoticity factors.}  
In the original work, chaoticity factors---extracted from correlation function measurements---were introduced as correction terms. These corrections primarily account for particle misidentification and weak-decay feed-down in correlation analyses. For our purposes, such corrections are unnecessary, since the proton and $\Lambda$ spectra used in this study have already been fully corrected for both misidentification and weak-decay feed-down. As demonstrated in Ref.~\cite{ALICE:2020ibs}, once these effects are removed, the proton chaoticity factor is unity, which is the value we adopt here.

For $\Lambda$ baryons, the situation is more subtle due to contributions from $\Sigma^0$ decays. All published $\Lambda$ spectra include feed-down from $\Sigma^0 \to \Lambda + \gamma$. Thermal model calculations (see Fig.~\ref{fig:n_over_p}) indicate that the $\Sigma^0$ fraction is approximately $27\%$ in Au+Au collisions, with only a weak dependence on collision energy. Because the proper decay length of the $\Sigma^0$ ($c\tau \approx 2.2\times 10^4$~fm) is much larger than the typical source size, $\Sigma^0$ baryons do not participate in the coalescence of $\Lambda$ hypernuclei (though they may contribute to the coalescence of hypothetical $\Sigma$ hypernuclei). Consequently, we scale the $\Lambda$ spectra by a factor of $1 - N_{\Sigma^0}/N_{\Lambda}$ (which is equivalent to the chaoticity), where the ratio $N_{\Sigma^0}/N_{\Lambda}$ is obtained from thermal model calculations.

\section{Nuclei wave functions}
\label{sec:wavefunction}

\subsection{Deuteron}

The default deuteron wave function employed in our study is the Argonne $v_{18}$ wave function~\cite{Wiringa:1994wb}. The Argonne $v_{18}$ interaction provides a high-precision fit to $N$–$N$ scattering data and the deuteron binding energy, and yields a deuteron wave function composed of both $S$- and $D$-wave components, thereby incorporating short-range repulsion and intermediate-range tensor correlations essential for describing the deuteron structure. For practical implementation, we adopt the Wigner density parameterization given in Ref.~\cite{Mahlein:2023fmx}. To assess model dependencies, we also consider alternative parameterizations based on the Hulthen~\cite{Heinz:1999rw} and Gaussian forms, as summarized in Ref.~\cite{Bellini:2020cbj}.

\subsection{Triton and ${}^{3}\rm{He}$}

The Gaussian wave function is employed for triton and ${}^{3}\rm{He}$. It has been shown in Ref.~\cite{Bellini:2020cbj} that coalescence calculations adopting this approximation yields good description of ${}^{3}\rm{He}$ yields measured at LHC energies. This could be due to the more compact nature of the triton and ${}^{3}\rm{He}$, compared to deuteron or the hypertriton. The Gaussian wave functions take the form: 

\begin{align}
\Phi_{t}(r_{nn}, r_{p}) &= \biggl(\frac{1}{3\pi^{2}b^{4}_{t}}\biggl)^{\frac{3}{4}}e^{-\frac{r_{nn}^{2}+\frac{4}{3}r_{p}^{2}}{4b_{b_{t}}^{2}}},\\
\Phi_{^{3}{\rm{He}}}(r_{pp}, r_{n}) &= \biggl(\frac{1}{3\pi^{2}b^{4}_{^{3}{\rm{He}}}}\biggl)^{\frac{3}{4}}e^{-\frac{r_{pp}^{2}+\frac{4}{3}r_{n}^{2}}{4b_{b_{^{3}{\rm{He}}}}^{2}}}. 
\end{align}

\noindent
Here, $r_{nn}$ and $r_{pp}$ denote the separations between two neutrons and two protons, respectively, while $r_{n}$ and $r_{p}$ represent the distances between the remaining nucleon and the corresponding pair. The parameter $b_{N}$ denotes the RMS charge radius, with values $b_{t} = 1.77~\text{fm}$ for the triton and $b_{{}^{3}\mathrm{He}} = 1.97~\text{fm}$ for ${}^{3}\mathrm{He}$~\cite{Ropke:2008qk}.

\subsection{Hypertriton ${}^{3}_{\Lambda}\rm{H}$}

The default wave function we employ is the Congleton wave function, defined in momentum space as: 
\begin{align} \label{eq:congleton}
\hat\Phi_{^{3}_{\Lambda}{\rm{H}}(d\Lambda)}(q)=A\frac{e^{-\frac{-q^2}{Q_{\Lambda}^2}}}{q^2+\alpha_{\Lambda}^2}. 
\end{align}

\begin{figure}[h!] 
    \centering
    \includegraphics[width=.9\linewidth]{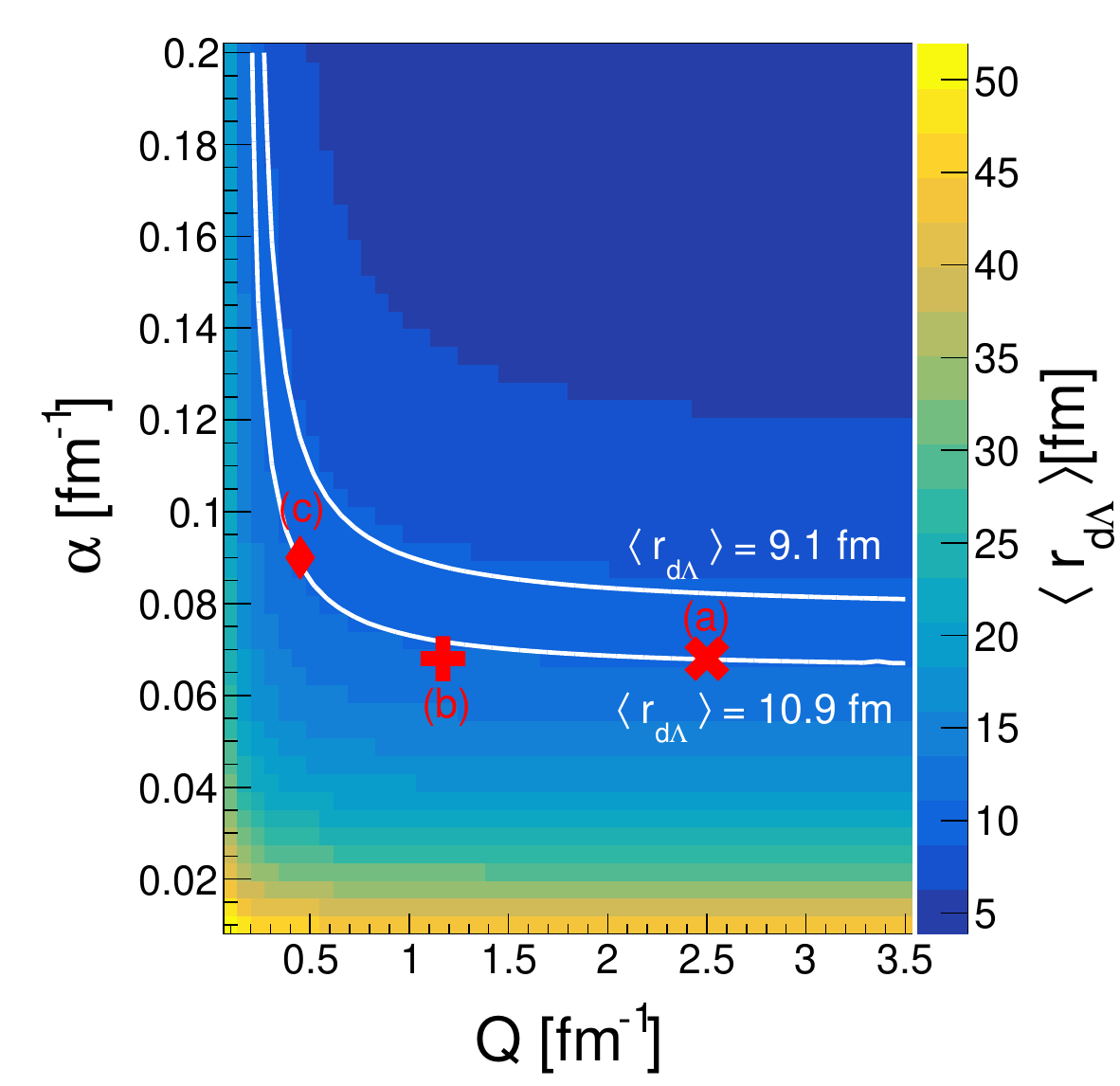} 
        \caption{The RMS distance between the deuteron and the $\Lambda$ ($\langle r_{d\Lambda} \rangle$) for the Congleton wave function using different paramterers $Q$ and $\alpha$. The white lines indicate the contours $\langle r_{d\Lambda} \rangle$=9.1 and 10.9 fm. The red markers represent the wave functions employed in this paper. }
    \label{fig:congleton}
\end{figure}

\begin{figure}[h!] 
    \centering
    \includegraphics[width=.9\linewidth]{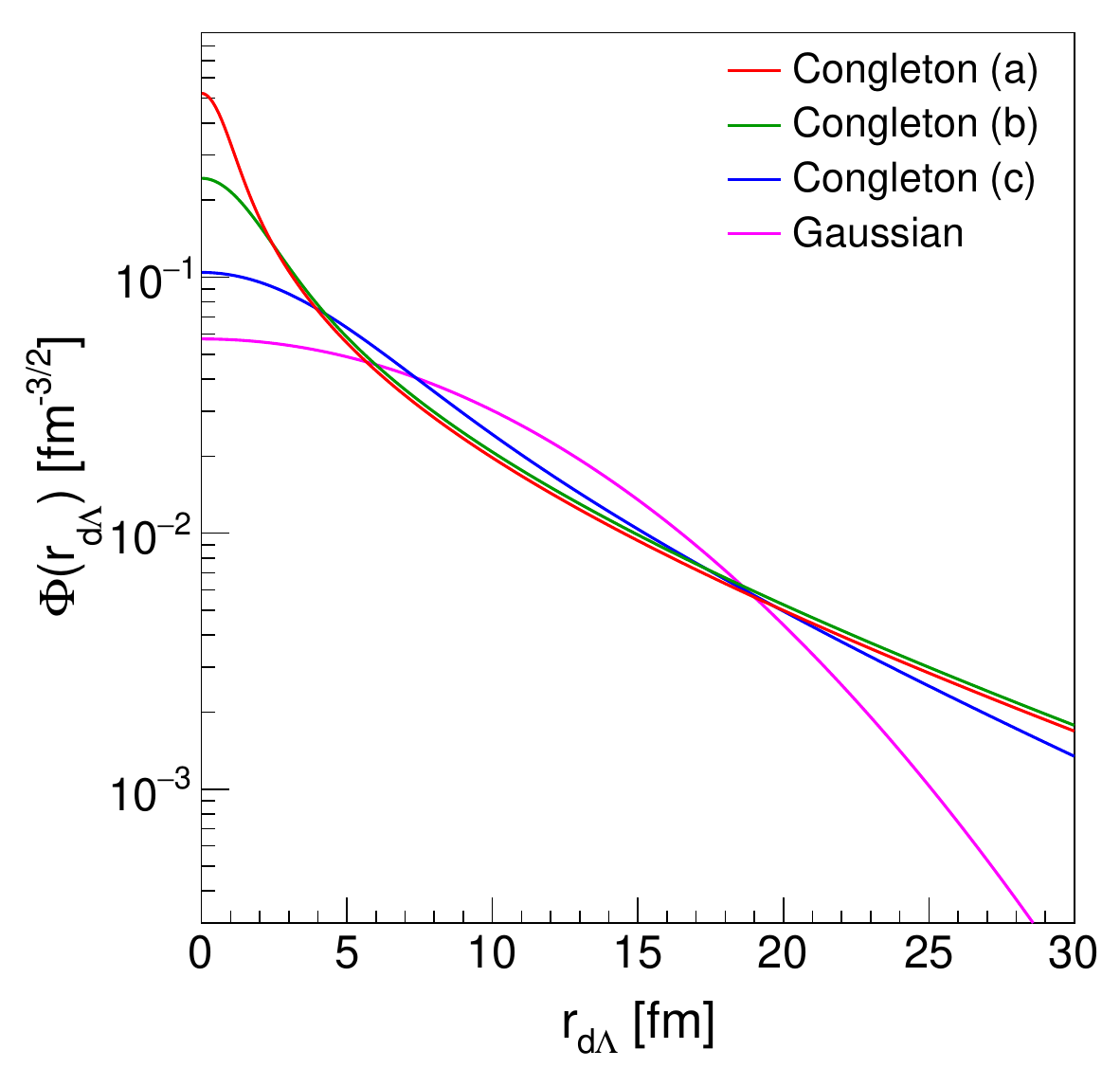} 
        \caption{The different ${}^{3}_{\Lambda}\rm{H}$ wave functions employed in this study. }
    \label{fig:congleton_wave_function}
\end{figure}

\noindent
Here, $A$ denotes a normalization constant. By modelling the hypertriton as a deuteron–$\Lambda$ bound state, Congleton obtained the parameters $Q_{\Lambda}=1.17~\text{fm}^{-1}$ and $\alpha_{\Lambda}=0.068~\text{fm}^{-1}$~\cite{Congleton:1992kk}. In addition to this parametrization, we also consider two others. The first is based on a three-body effective field theory (EFT) calculation from Ref.~\cite{Hildenbrand:2019sgp}. It has been shown in Ref.~\cite{Bellini:2020cbj} that the Congleton wave function, with modified parameters $Q_{\Lambda}=2.5~\text{fm}^{-1}$ and $\alpha_{\Lambda}=0.068~\text{fm}^{-1}$, reproduces the EFT results well. The RMS deuteron–$\Lambda$ separation $\langle r_{d\Lambda} \rangle$ corresponding to both of these parametrizations is approximately $10.8~\text{fm}$, consistent with the value $9.8^{+1.1}_{-0.7}~\text{fm}$ inferred from the the ${}^{3}_{\Lambda}\mathrm{H}$ binding energy $0.163\pm0.036$~MeV~\cite{Liu:2024ygk, Eckert:2022dyz, Kasagi:2025mvh, ALICE:2022sco, STAR:2019wjm, Chaudhari1968, Juric:1973zq, Mayeur1966ADO, ammar1962, Prakash1961OnTB}. Since the deuteron-$\Lambda$ interaction is not yet well constrained by experiments, we also select another set of parameters that lie close to the $\langle r_{d\Lambda}\rangle =10.9$~fm contour, as shown in Fig.~\ref{fig:congleton}. A summary of the parameters used is given in Tab.~\ref{tab:congparams}. The wave function of the $p$–$n$ subsystem is modeled by a Gaussian with $b_{pn} = 1.52~\text{fm}$, as detailed in Ref.~\cite{Bellini:2020cbj}. In addition to the Congleton wave function, we will also employ a Gaussian ansatz for comparison:

\begin{align} \label{eq:gaussianhypertriton}
\Phi_{{}^{3}_{\Lambda}\mathrm{H}}(r_{pn}, r_{\Lambda}) 
&= \left(\frac{1}{3\pi^{2}b_{pn}^{2}b_{\Lambda}^{2}}\right)^{\tfrac{3}{4}}
   e^{  -\frac{r_{pn}^{2}}{4b_{pn}^{2}} - \frac{r_{\Lambda}^{2}}{3b_{\Lambda}^{2}}}, 
\end{align}
\noindent 
with $b_{\Lambda}=7.2$ fm, which also corresponds to $\langle r_{d\Lambda} \rangle=10.8$~fm. The wave functions employed and their radial probability distributions in coordinate space are shown in Fig.~\ref{fig:congleton_wave_function} and~\ref{fig:rad_prob_congleton_wave_function} respectively. We note here that all four wave functions have $\langle r_{d\Lambda} \rangle \approx 10.8$ fm, which corresponds to the lower limit of the measured ${}^{3}_{\Lambda}\mathrm{H}$ binding energy. Our motivation is to investigate the effect arising from the shape of the wave function while keeping the average spatial extent $\langle r_{d\Lambda} \rangle$ constant. In Appendix~\ref{sec:appendix_2}, we will instead choose four different wave functions with $\langle r_{d\Lambda} \rangle \approx 9.1$ fm, corresponding to the upper limit of the measured ${}^{3}_{\Lambda}\mathrm{H}$ binding energy, to quantify the effect arising from the uncertainty associated to the binding energy. 

\begin{table}[htbp]
\centering
\caption{Summary of Congleton $^{3}_{\Lambda}\rm{H}$ wave function parametrizations considered in this work.}
\label{tab:congparams}
\begin{tabular}{lccc}
\hline\hline
 & $Q_{\Lambda}$ (fm$^{-1}$) & $\alpha_{\Lambda}$ (fm$^{-1}$) & Parameterization \\
\hline
(a) & 2.5 & 0.068 & 3-body EFT~\cite{Hildenbrand:2019sgp} \\
(b) & 1.17 & 0.068 & 2-body closure approx.~\cite{Congleton:1992kk}\\
(c) & 0.45 & 0.090 & - \\
\hline\hline
\end{tabular}
\end{table}

\section{Data-driven source size parameterization}
\label{sec:source}

The conventional method for extracting the source radius is through correlation function measurements~\cite{PhysRevLett.3.181, PhysRev.120.300}, analogous to the determination of stellar angular diameters via astrophysical intensity correlation measurements by R.~Hanbury Brown and R.~Q.~Twiss (HBT)~\cite{HanburyBrown:1956bqd}. Such measurements have been extensively performed at LHC energies~\cite{ALICE:2025wuy, ALICE:2020ibs}. In contrast, for Au+Au collisions at $\sqrt{s_{\mathrm{NN}}}=3$--200~GeV, results on baryon correlations are still limited. 

To overcome this limitation, we propose using the deuteron coalescence parameter $B_{2}$, which can be obtained from published deuteron and proton spectra via Eq.~\ref{eq:bn}. The measured $B_{2}$, when combined with Eq.~\ref{eq:b2}, provides an estimate of the source radius $R_{\rm inv}$. It should be emphasized that predicting deuteron yields from a source radius extracted in this manner is trivial, since they are by construction consistent with the input measurements. However, by assuming that the source radius remains unchanged for nuclei with $A>2$, this approach enables predictions of heavier nuclei yields. In particular, it allows for an estimate of ${}^{3}_{\Lambda}\rm{H}$ production, which is the primary focus of this study.




\subsection{Validation using Data from $\sqrt{s_{\rm{NN}}}$=5.02 TeV Pb+Pb Collisions}

Before applying our methodology to Au+Au collisions, we first validate it using data from Pb+Pb collisions at $\sqrt{s_{\rm NN}}=5.02$~TeV measured by the ALICE Collaboration. At this energy, ALICE has extracted the source size as a function of transverse momentum $p_{T}$ via correlation function analyses~\cite{ALICE:2025wuy}. This provides an opportunity to benchmark our approach by comparing the source radii obtained from our method, based on the deuteron coalescence parameter $B_{2}$, with those from correlation functions. 

\begin{figure}[h!] 
    \centering
    \includegraphics[width=.85\linewidth]{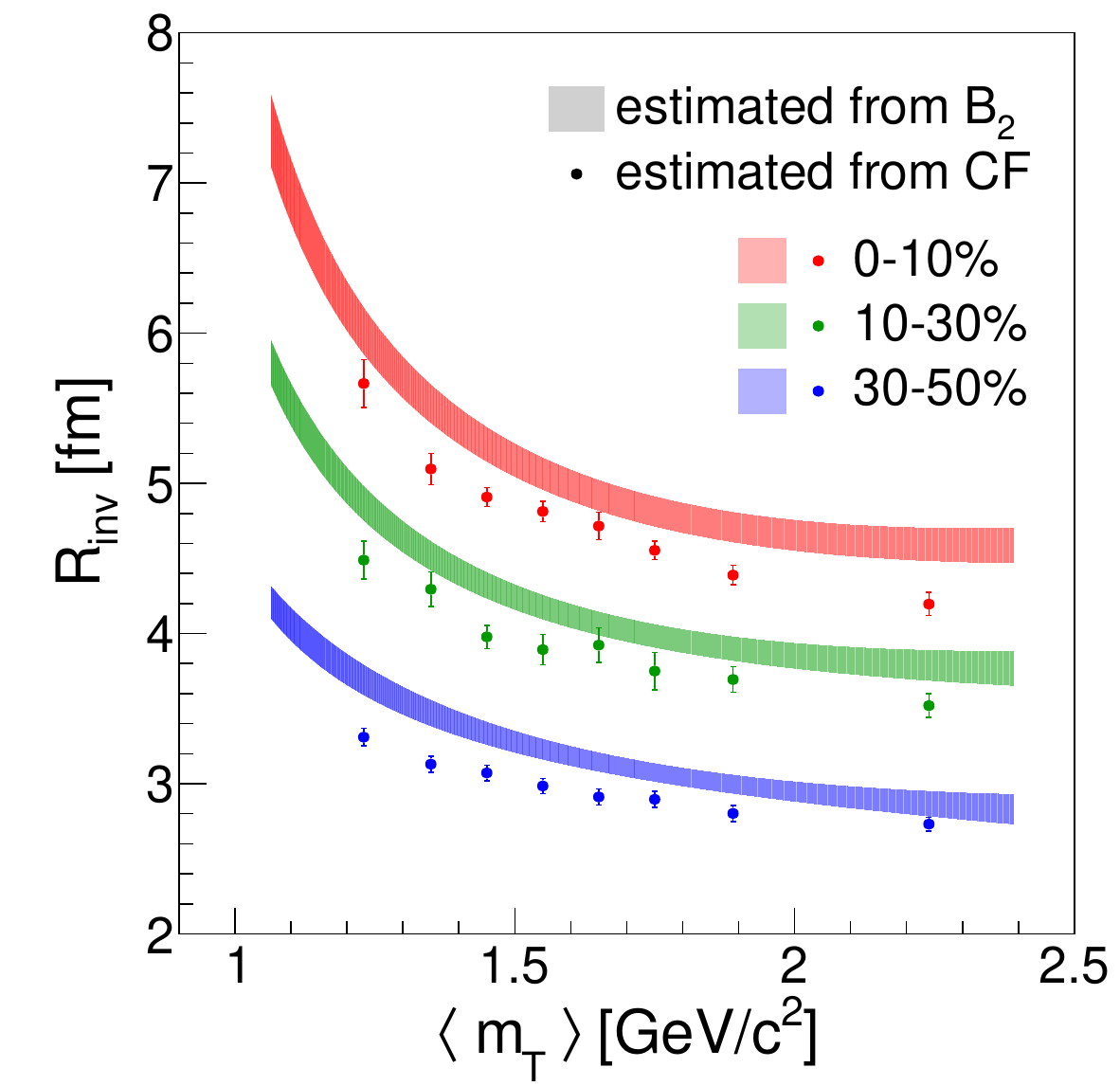} 
        \caption{Comparison of the nucleon source size extracted from deuteron $B_{2}$ data~\cite{ALICE:2022veq} (colored bands) and proton-proton correlation functions~\cite{ALICE:2025wuy} (dashed lines).}
    \label{fig:pbpb5p02}
\end{figure}

Figure~\ref{fig:pbpb5p02} presents the source radii extracted by the two methods as functions of the mean transverse mass, for three centrality classes. The shaded bands represent the propagated uncertainties arising from the measured $B_{2}$~\cite{ALICE:2022veq}. The comparison shows that, while the two methods exhibit very similar trends across all centralities and transverse masses, the radii extracted from $B_{2}$ are consistently larger than those from correlation functions by an approximately constant factor of $\sim 8\%$.

The origin of this discrepancy is not fully understood and may stem from approximations inherent in either approach. We interpret this difference as a systematic uncertainty in the source size extraction. Unless otherwise stated, in the remainder of this work we scale down the source size extracted from $B_{2}$ by $4\%$ and assign a systematic uncertainty of $\pm 4\%$.

\subsection{Energy Dependence of Source Radius}

The deuteron and proton spectra measured by the STAR Collaboration in Au+Au collisions over $\sqrt{s_{\rm{NN}}}=3$--200~GeV~\cite{STAR:2019sjh, STAR:2022hbp, STAR:2023uxk} enable the calculation of $B_{2}$ differentially as a function of $p_{T}$ and centrality. However, the measured spectra only cover the region $p_{T}/A \gtrsim 0.5$~GeV$/c$ due to limited experimental acceptance, and thus extrapolation is required. To this end, we extrapolate the proton and deuteron spectra to the unmeasured region using the blast-wave parameterization~\cite{Schnedermann:1993ws}:

\begin{align}
\frac{1}{2\pi p_{T}} \frac{d^{2}N}{dp_{T}dy} \propto 
\int_{0}^{R} r\, dr\, m_{T}\,
I_{0}\!\left(\frac{p_{T}\sinh\rho(r)}{T_{\rm kin}}\right) \nonumber \\
\times \, K_{1}\!\left(\frac{m_{T}\cosh\rho(r)}{T_{\rm kin}}\right),
\label{eq:blast}
\end{align}

\noindent where $m_{T}$ is the transverse mass of the particle, $I_{0}$ and $K_{1}$ are modified Bessel functions, and $\rho(r) = \tanh^{-1}(\beta_{T})$ is the boost angle associated with the transverse flow velocity $\beta_{T}$. For $0 \le r \le R$, the radial flow velocity profile is parametrized as $\beta_{T}(r) = \beta_{S} \left(\frac{r}{R}\right)^{n}$, where $\beta_{S}$ is the surface velocity and $n$ characterizes the shape of the velocity profile. In this study, $n$ is fixed to unity for Au+Au collisions at $\sqrt{s_{\rm{NN}}}=3$--200~GeV, while it is treated as a free parameter at LHC energies.

\begin{figure}[h!] 
    \centering
    \includegraphics[width=.49\linewidth]{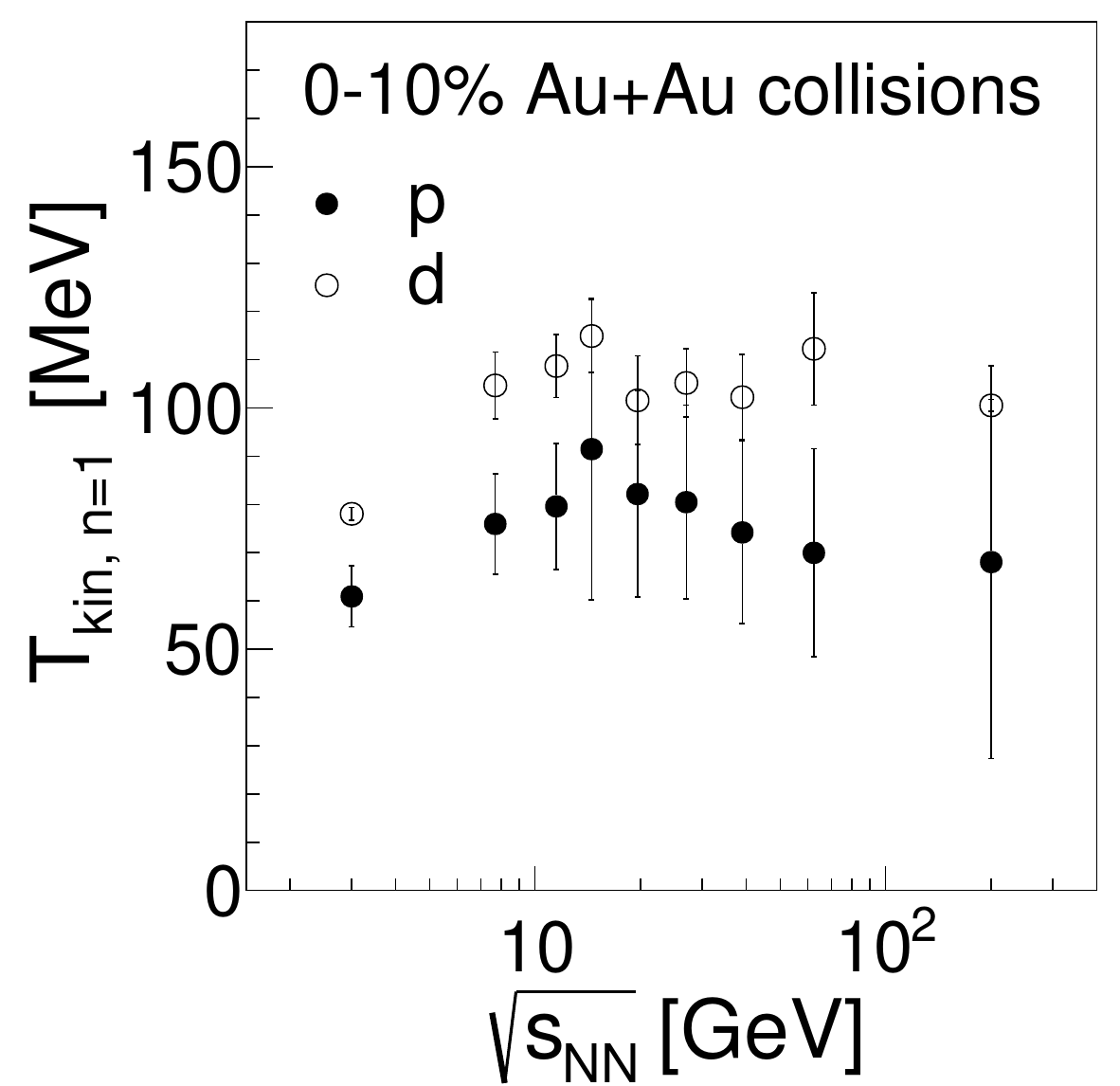} 
    \includegraphics[width=.49\linewidth]{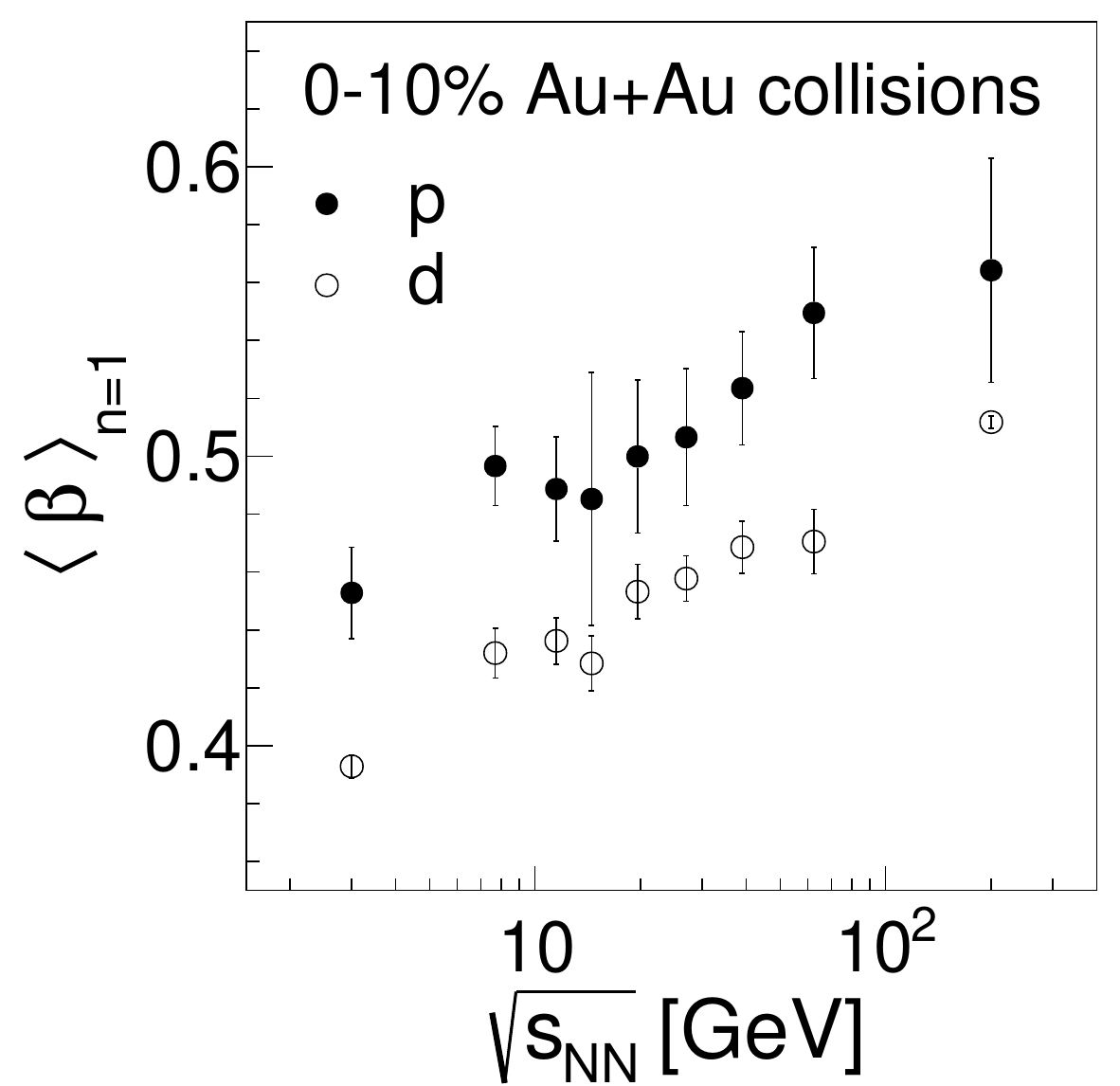} 
        \caption{Blast-wave parameters $T_{\rm{kin}}$ and $\beta$ for protons and deuterons obtained by fitting to $0-10\%$ their respective $p_{T}$ spectra from \snn=3 to 200 GeV Au+Au collisions~\cite{STAR:2017sal, STAR:2023uxk, STAR:2019sjh}. The flow profile $n$ is fixed to one in these fits.}
    \label{fig:freezeout}
\end{figure}

The fitted kinetic freeze-out temperature and surface velocity for $0$--$10\%$ central Au+Au collisions are shown in Fig.~\ref{fig:freezeout}. A systematic difference is observed between the deuteron and proton freeze-out parameters: deuterons exhibit consistently higher kinetic freeze-out temperatures and lower radial flow velocities compared to protons. This suggests that protons and deuterons do not undergo identical hydrodynamic evolution, consistent with the expectation that deuterons are predominantly formed through the coalescence mechanism. 

It should be emphasized that such extrapolations introduce additional uncertainties in the unmeasured low-$p_{T}/A$ region ($<0.5$~GeV$/c$). Consequently, extra systematic uncertainties must be assigned when computing integrated quantities, as will be discussed in later sections.

\begin{figure}[h!] 
    \centering
    \includegraphics[width=.85\linewidth]{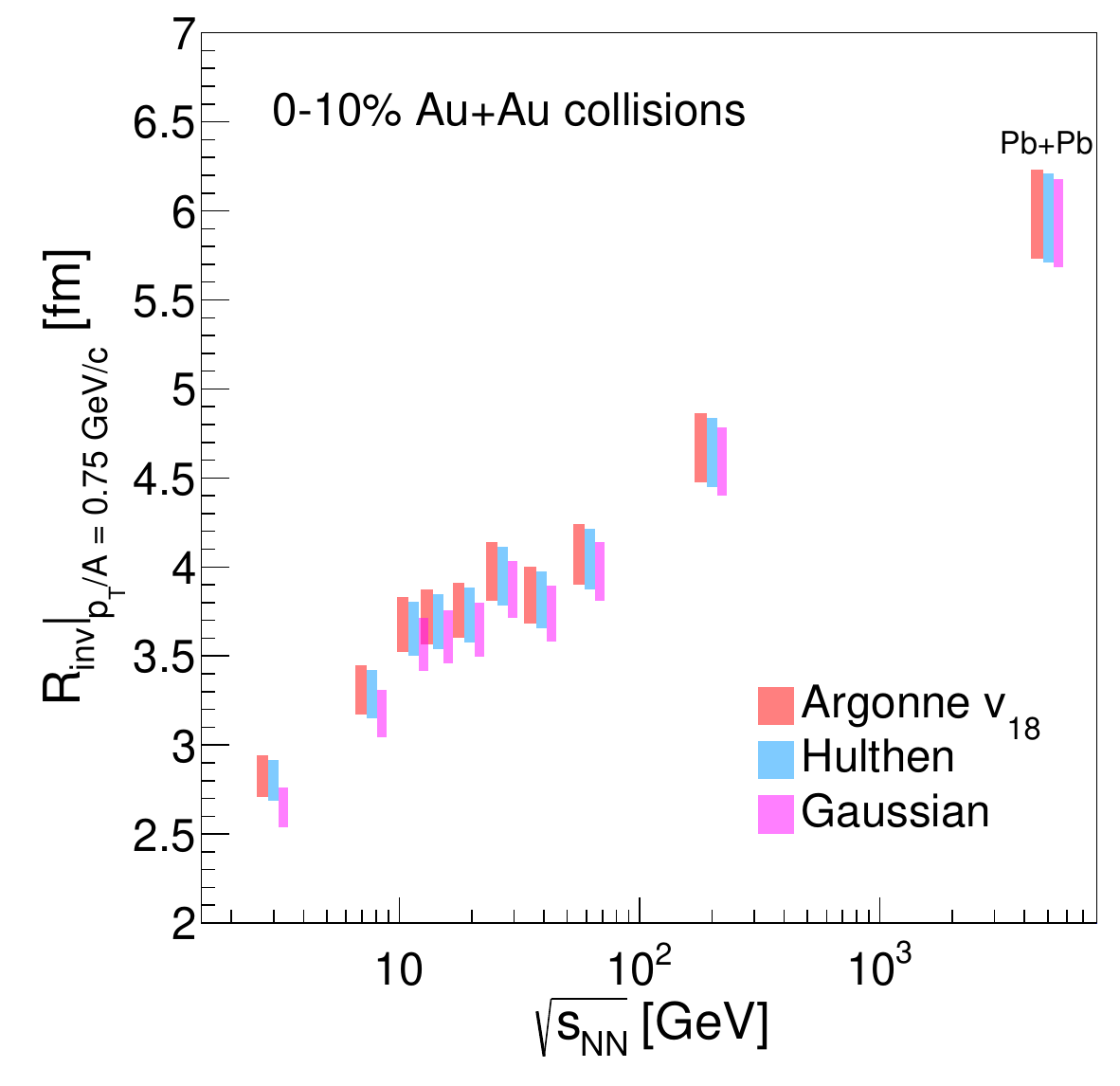} 
        \caption{Nucleon source size extracted from deuteron $B_{2}$ data~\cite{STAR:2019sjh, STAR:2022hbp, STAR:2023uxk, ALICE:2022veq, ALICE:2019hno} using coalescence calculations with different deuteron wave functions.}
    \label{fig:rinv_vs_snn}
\end{figure}

Fig.~\ref{fig:rinv_vs_snn} shows the source radius $R_{\rm{inv}}$ obtained from deuteron~\cite{STAR:2023uxk, ALICE:2022veq, STAR:2019sjh} and proton~\cite{STAR:2017sal, STAR:2023uxk, ALICE:2019hno} spectra as a function of collision energy $\sqrt{s_{\rm{NN}}}$ at $p_{T}/A=0.75$ GeV$/c$. As discussed in the previous section, an uncertainty in the source size, in which the upper edge corresponds to the result using Eq.~\ref{eq:b2}, while the lower edge is $8\%$ lower is implemented. Unless otherwise stated, this uncertainty will be applied to all calculations shown. In addition to the Argonne $v_{18}$ wave function, the Hulthen and Guassian wave functions are employed. The results are very similar between the different wave functions, leading to less to $1\%$ difference at \snn=5.02 TeV. At the lowest energy \snn=3 GeV, the differences are slightly larger, but still less than $6\%$. 

The extracted source size increases from \snn=3 GeV to around 11.5 GeV, and stays level until 62.4 GeV. Beyond 62.4 GeV, it rises steadily. This energy dependence is qualitatively same as that of the radial flow velocity extracted from blast-wave fits to spectra~\cite{STAR:2017sal}; i.e. a system with larger flow velocity will lead to an increase in source radius. 

\subsection{Charged-Particle Multiplicity Dependence of Source Radius}

\begin{figure}[h!] 
    \centering
    \includegraphics[width=.95\linewidth]{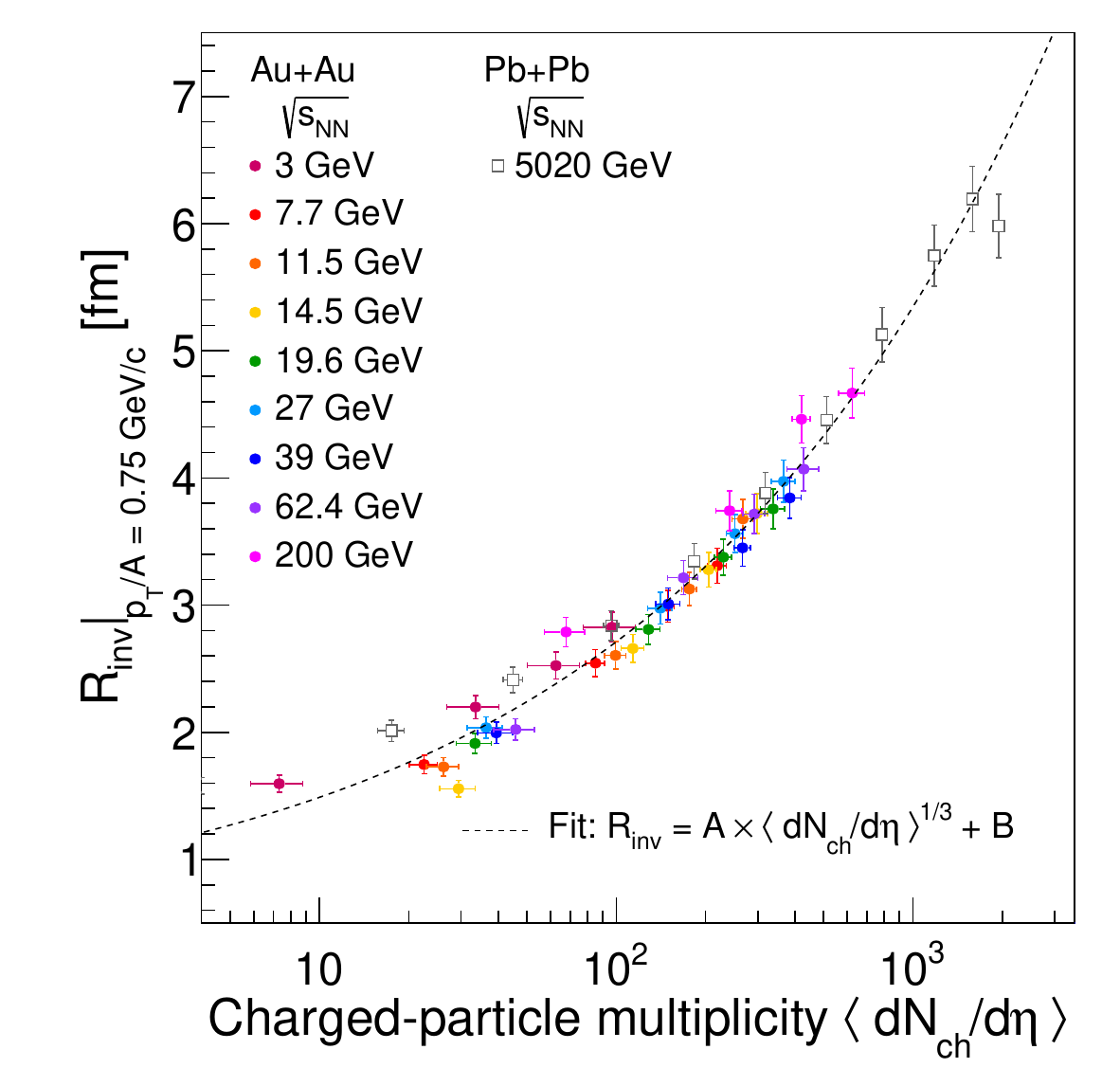} 
        \caption{Nucleon source size extracted from deuteron $B_{2}$ data~\cite{STAR:2019sjh, STAR:2022hbp, STAR:2023uxk, ALICE:2022veq, ALICE:2019hno} using coalescence calculations with Argonne $v_{18}$ wave function as a function of charged particle multiplicity. The dotted line represents a fit using the function $R_{\rm{inv}}= A \times  \langle dN_{\rm{ch}}/d\eta \rangle +B$. }
    \label{fig:rinv_vs_nch}
\end{figure}

It is instructive to investigate the source radius as a function of the average charged-particle multiplicity $\langle dN_{\rm{ch}}/d\eta \rangle$, since multiplicity is closely related to the size of the created system~\cite{Lisa:2005dd}. In particular, it has been shown that in Pb+Pb collisions at $\sqrt{s_{\mathrm{NN}}}=5.02$~TeV, the source radius extracted from correlation functions can be well described by the functional form $R = A \times \langle dN_{\rm{ch}}/d\eta \rangle ^{1/3} + B$~\cite{ALICE:2025wuy}. 

To estimate the charged-particle multiplicity in Au+Au collisions, the measured yields of $\pi^{\pm}$, $K^{\pm}$, and $p(\bar{p})$~\cite{STAR:2017sal} at \snn=7.7--200 GeV are combined with a stochastic approach~\cite{Ide:2003xv}. For \snn=3 GeV, since $\pi^{\pm}$ data are not available, we instead use the transport model Ultrarelativistic Quantum Molecular Dynamics (UrQMD)~\cite{Bleicher:1999xi} to estimate the charged-particle multiplicity. Figure~\ref{fig:rinv_vs_nch} shows the source radius at $p_{T}/A=0.75$~GeV$/c$ as a function of charged-particle multiplicity. The results are obtained using deuteron and proton yields from four centrality classes ($0$--$10\%$, $10$--$20\%$, $20$--$40\%$, $40$--$80\%$) in Au+Au collisions and ten centrality classes ($0$--$5\%$, $5$--$10\%$, $10$--$20\%$, $20$--$30\%$, $30$--$40\%$, $40$--$50\%$, $50$--$60\%$, $60$--$70\%$, $70$--$80\%$, $80$--$90\%$) in Pb+Pb collisions. 

A global fit of all data to the functional form $R = A \times \langle dN_{\rm{ch}}/d\eta \rangle ^{1/3} + B$ yields $A = 0.501 \pm 0.011$ and $B = 0.36 \pm 0.06$. The fit quality, with $\chi^2/\mathrm{NDF} = 62/40$, indicates that this scaling relation describes the data quite well, independent of collision energy or system. This result suggests that the coalescence probability depends primarily on the system size, rather than the collision energy. A small deviation is observed between Pb+Pb collisions at $\sqrt{s_{\mathrm{NN}}}=5.02$~TeV and Au+Au collisions at $\sqrt{s_{\mathrm{NN}}}=7.7$--$62.4$~GeV for multiplicities below $\langle N_{\rm ch} \rangle \lesssim 60$, which may warrant further investigation in the future. 

\begin{figure}[h!] 
    \centering
    \includegraphics[width=.99\linewidth]{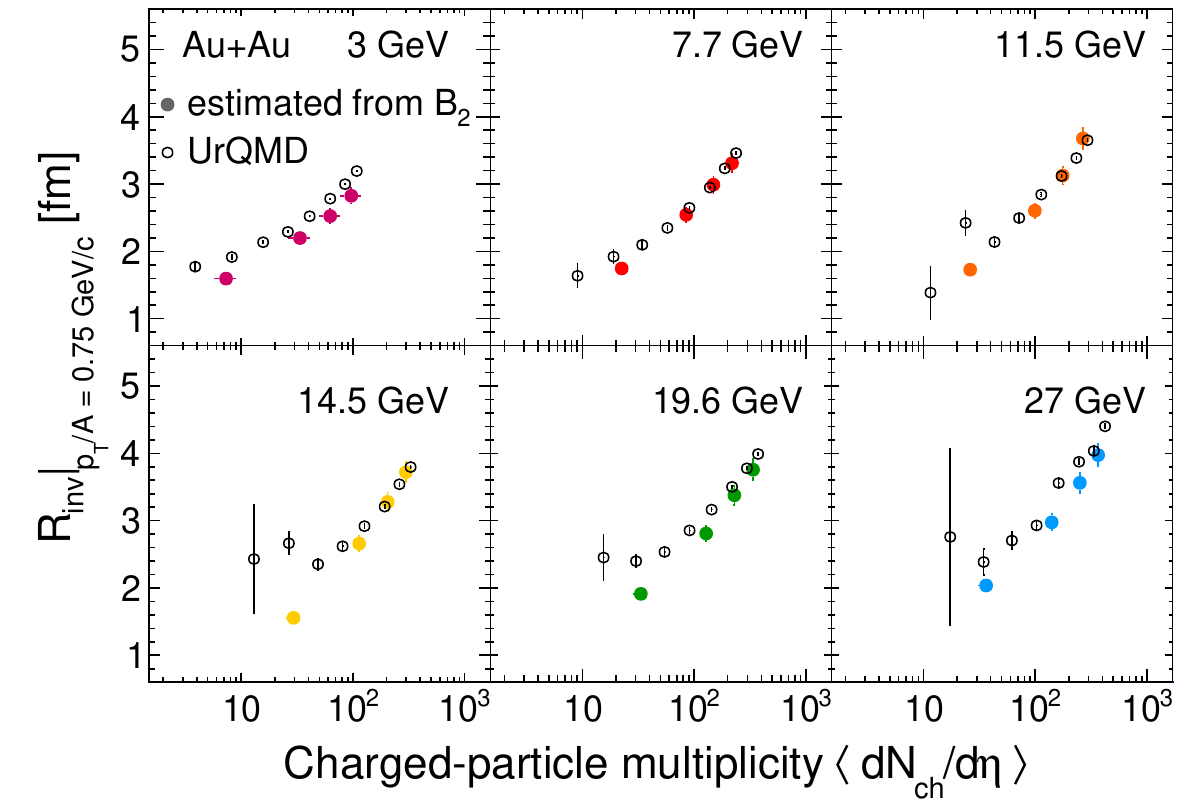} 
        \caption{Nucleon source size at $p_{T}/A=0.75$ GeV$/c$ extracted from deuteron $B_{2}$ data~\cite{STAR:2019sjh, STAR:2022hbp, STAR:2023uxk} using coalescence calculations with Argonne $v_{18}$ wave function (solid dots) and from UrQMD~\cite{Bleicher:1999xi} (open dots), as a function of charged particle multiplicity. }
    \label{fig:rinv_urqmd}
\end{figure}

We further compare our estimated source radii with results from the transport model UrQMD (v4.0) in cascade mode. For each event, proton pairs with relative momentum in the pair rest frame $k^* < 100$~MeV$/c$ are selected for analysis. A Gaussian fit to the pair distribution is then used to extract the source radius. In Fig.~\ref{fig:rinv_urqmd}, we present the comparison at $|y|<0.5$ and $p_{T}/A=0.75$ GeV$/c$ between the radii obtained from UrQMD and those inferred from the $B_{2}$ data over the range $\sqrt{s_{\rm{NN}}}=3$--27~GeV. The two methods show good agreement for $\sqrt{s_{\rm NN}} = 7.7$--27~GeV, except possibly in the most peripheral collisions, thereby lending confidence to our approach. At $\sqrt{s_{\rm{NN}}}=3$~GeV, a deviation of $\sim10\%$ is observed: UrQMD systematically predicts slightly larger radii across all centralities. This discrepancy may, in part, arise from the sensitivity of the source radius to the EoS at low collision energies.
Future precise femtoscopic measurements of the source size at these energies are crucial. 

With the source radius and the wave function specified as above, we can compute $B_{3}$ for triton, $^{3}\mathrm{He}$, and ${}^{3}_{\Lambda}\mathrm{H}$ according to Eq.~\ref{eq:b3} and~\ref{eq:b3lambda}. Combining $B_{3}$ with the measured proton and $\Lambda$ spectra enables predictions for the $p_{T}$ spectra of triton, $^{3}\mathrm{He}$, and ${}^{3}_{\Lambda}\mathrm{H}$. From these, we can also obtain integrated observables such as the $p_{T}$-integrated yields and the mean transverse momentum. It should be noted, however, that the estimation of the source size at low $p_{T}$ relies on extrapolation, which introduces uncertainties, as illustrated in Fig.~\ref{fig:freezeout}. Performing estimations of the uncertainty of such extrapolations separately for each collision energy is not straightforward. Instead, we adopt a simplified treatment: experimental studies based on variations of fit functions have quantified the associated uncertainty to be approximately $10\%$ for $p_{T}$-integrated yields and $4\%$ for the mean transverse momentum~\cite{STAR:2022hbp}. These values will be used when quoting uncertainties for all $p_{T}$-integrated predictions and will be added in quadrature to the aforementiond $4\%$ uncertainty associated to the uncertainty in the source size. For ratios such as $S_{3} = ({}^{3}_{\Lambda}\mathrm{H}/\Lambda)/({}^{3}\mathrm{He}/p)$, where the predictions for the integrated yields appear in both denominator and numerator, we assume that these uncertainties  cancel. 

\section{Results}
\label{sec:results}

\subsection{Deuteron}
\subsubsection{Transverse Momentum Spectra}

\begin{figure}[h!] 
    \centering
    \includegraphics[width=.99\linewidth]{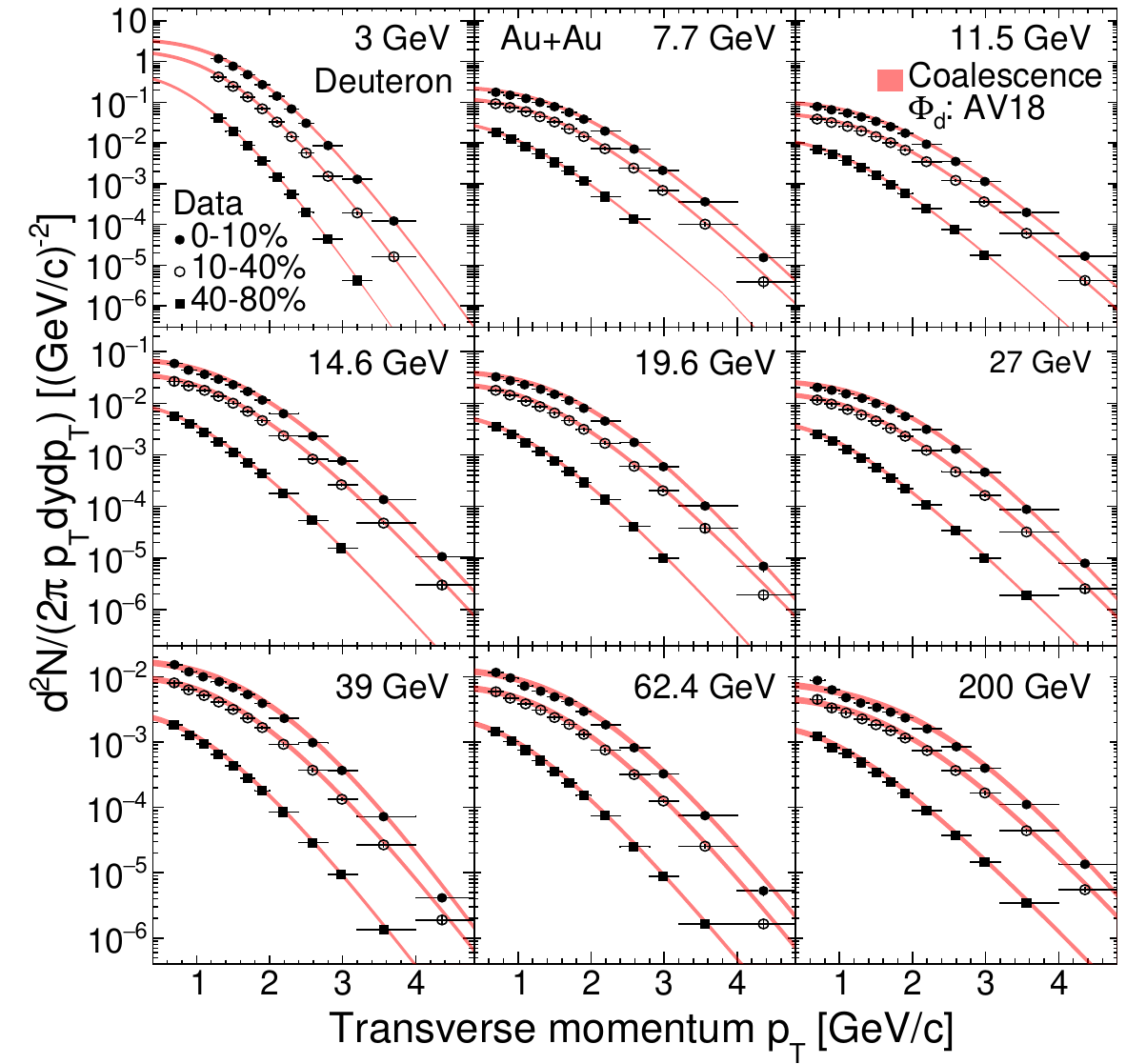} 
    \caption{Deuteron transverse momentum spectra at \snn=3, 7.7, 11.5, 14.6, 19.6, 27, 39, 62.4, and 200 GeV. The data~\cite{STAR:2019sjh, STAR:2023uxk} in three different centralities is compared with coalescence calculations. The Argonne $v_{18}$ deuteron wave function is used in the source size calculation.}
    \label{fig:deuteron_spectra_coal}
\end{figure}

Figure~\ref{fig:deuteron_spectra_coal} presents the transverse momentum ($p_{T}$) spectra of deuterons at \snn = 3, 7.7, 11.5, 14.6, 19.6, 27, 39, 62.4, and 200 GeV. The experimental data~\cite{STAR:2019sjh, STAR:2023uxk} for three centrality classes, $0-10\%$, $10-40\%$, and $40-80\%$ are compared with coalescence model calculations, shown using the red lines. In the source size determination, the Argonne $v_{18}$ deuteron wave function is employed. The width of the red line represents an $8\%$ difference in the source size, as discussed in the earlier section. The coalescence calculations, by construction, is consistent with the data. 

\subsection{Triton}
\subsubsection{Transverse Momentum Spectra}

\begin{figure}[h!] 
    \centering
    \includegraphics[width=.99\linewidth]{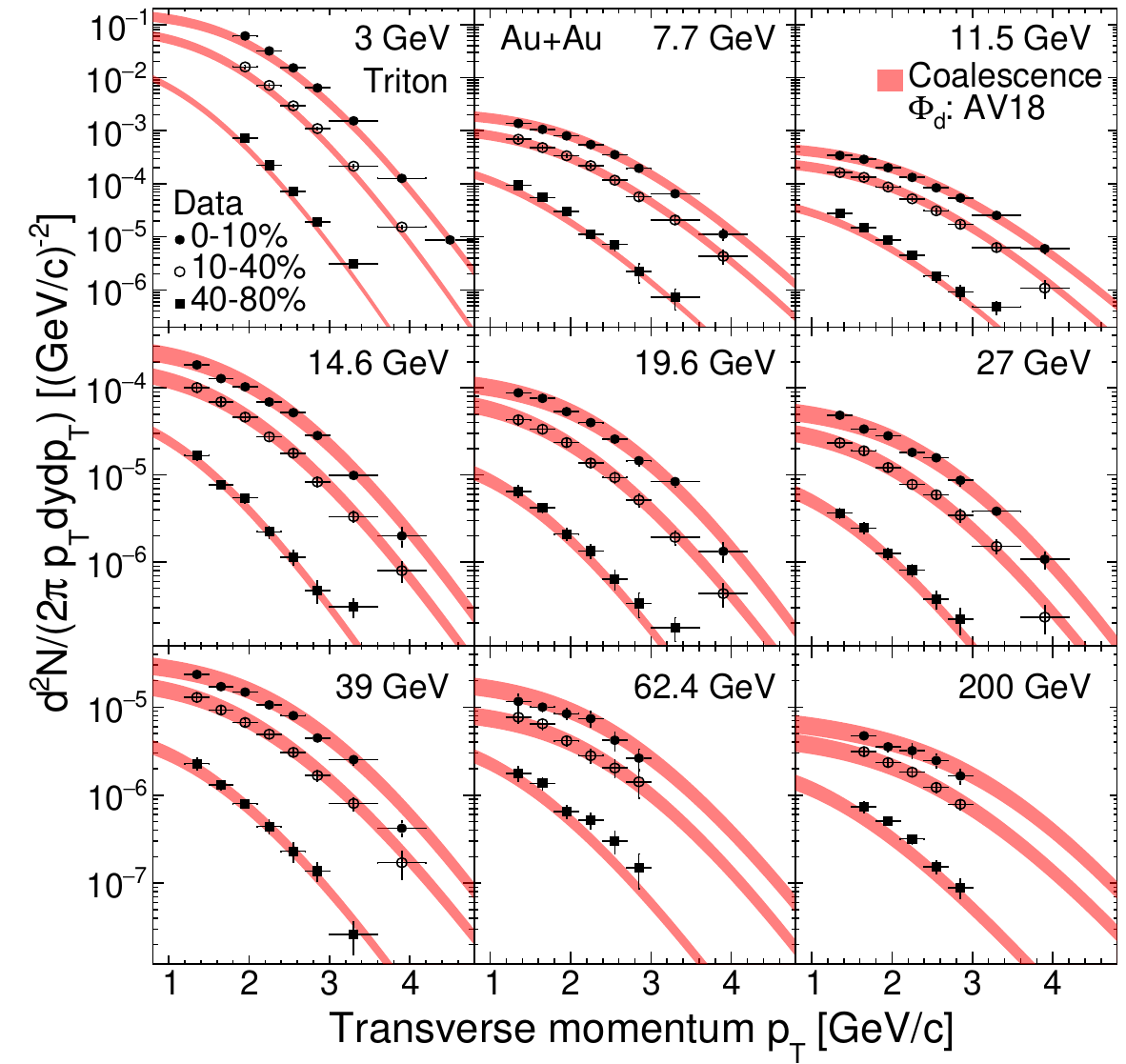} 
    \caption{Triton transverse momentum spectra at \snn=3, 7.7, 11.5, 14.6, 19.6, 27, 39, 62.4, and 200 GeV. The data~\cite{STAR:2022hbp} in three different centralities is compared with coalescence calculations. The Argonne $v_{18}$ deuteron wave function is used in the source size calculation.}
    \label{fig:triton_spectra_coal}
\end{figure}

Figure~\ref{fig:triton_spectra_coal} shows the transverse momentum ($p_{T}$) spectra of tritons at \snn = 3, 7.7, 11.5, 14.6, 19.6, 27, 39, 62.4, and 200 GeV. Experimental data~\cite{STAR:2022hbp} for three centrality classes ($0$–$10\%$, $10$–$40\%$, and $40$–$80\%$) are compared with coalescence model calculations. In determining the source size, the Argonne $v_{18}$ deuteron wave function is used. The width of the red line corresponds to an $8\%$ variation in the source size, as discussed earlier. The broader band compared to deuterons indicates a stronger sensitivity to the source size.

\begin{figure*}[t] 
    \centering
    \includegraphics[width=.6\linewidth]{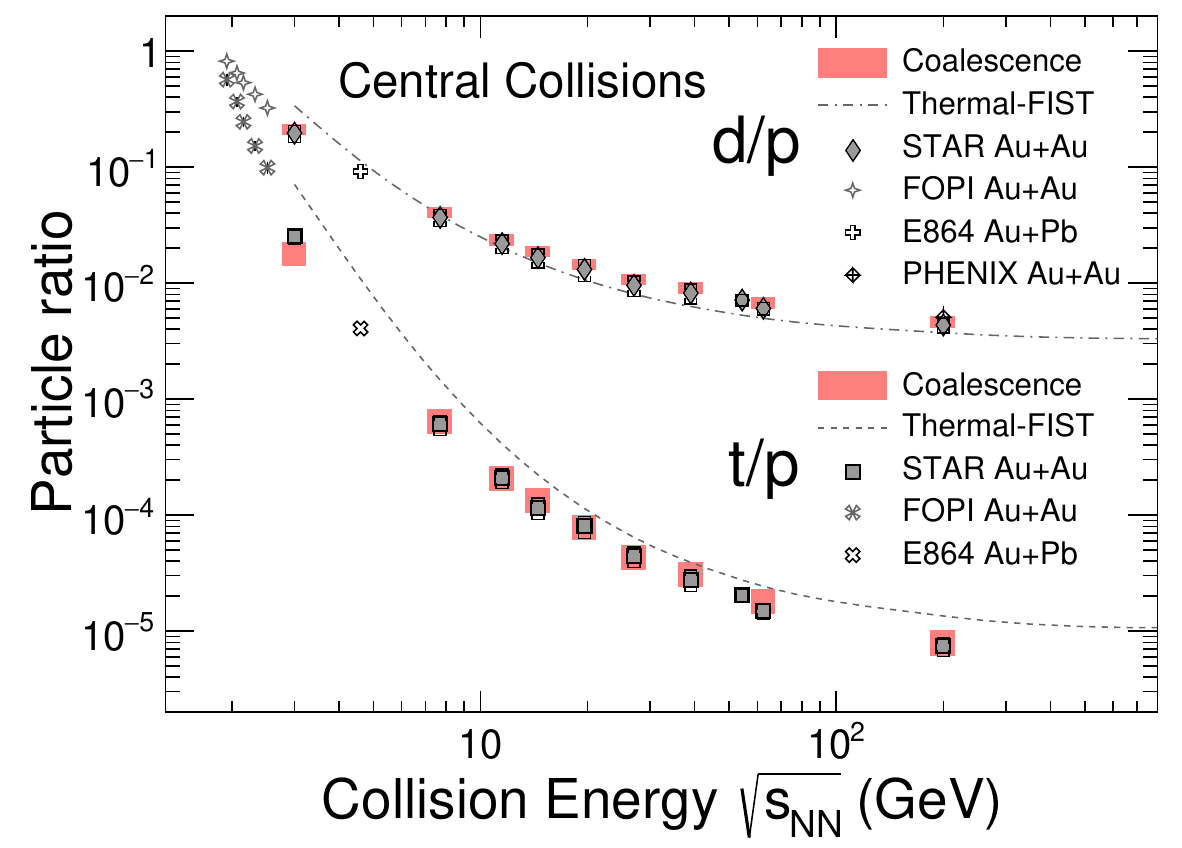} 
        \caption{Yield ratios $N_{d}/N_{p}$ and $N_{t}/N_{p}$ in central heavy-ion collisions as a function of $\sqrt{s_{\rm{NN}}}$. The data from FOPI~\cite{FOPI:2006ifg}, E864~\cite{E864:2000auv}, PHENIX~\cite{PHENIX:2004vqi, PHENIX:2003iij}, and STAR~\cite{STAR:2019sjh, STAR:2023uxk, STAR:2022hbp} are compared with coalescence (red bands) and Thermal-FIST~\cite{Vovchenko:2019pjl} (dashed lines) calculations. }
    \label{fig:dptp_vs_snn}
\end{figure*}

Overall, the coalescence calculations reproduce the data well. The best agreement is observed in central collisions, while a slight underestimation appears in peripheral collisions, which may be due to approximations in the coalescence model that become more significant in smaller systems. For comparison, calculations with source sizes obtained using Hulthen and Gaussian wave functions are presented in Appendix~\ref{sec:appendix_1}. The Hulthen results agree closely with the Argonne $v_{18}$ case, as expected from their nearly identical source sizes. In contrast, the Gaussian wave function leads to significant deviations, particularly in $40$–$80\%$ peripheral collisions, where the calculated spectra exceed the data by up to a factor of two at \snn = 3–39 GeV. This highlights the importance of employing realistic wave functions in coalescence calculations.

\subsubsection{$p_{T}$-Integrated Nuclei-to-Nucleon Yield Ratios}

Figure~\ref{fig:dptp_vs_snn} shows the yield ratios $N_{d}/N_{p}$ and $N_{t}/N_{p}$ in $0$–$10\%$ Au+Au and Au+Pb collisions as a function of $\sqrt{s_{\rm NN}}$, compared with predictions from coalescence and thermal models. The thermal results are obtained with the Thermal-FIST package~\cite{Vovchenko:2019pjl}, using freeze-out parameters $(T,\mu_B)$ from Ref.~\cite{Vovchenko:2015idt} and including feed-down contributions from unstable nuclei~\cite{Vovchenko:2020dmv}. Data from FOPI~\cite{FOPI:2006ifg}, E864~\cite{E864:2000auv}, PHENIX~\cite{PHENIX:2004vqi, PHENIX:2003iij}, and STAR~\cite{STAR:2019sjh, STAR:2023uxk, STAR:2022hbp} are also shown. Both models reproduce the $N_{d}/N_{p}$ ratio. This is expected for the coalescence model, since the coalescence source size is constrained by deuteron and proton data. For $N_{t}/N_{p}$, however, the thermal model overshoots the data by about a factor of two, while the coalescence calculations describe the measurements over a broad energy range, $\sqrt{s_{\rm NN}}=3$–200 GeV. As discussed in the previous section, we assign an uncertainty to the source size. This uncertainty translates into uncertainties of approximately $20\%$ and $30\%$ in the absolute yields of $N_d$ and $N_t$, respectively, and consequently in the ratios $N_d/N_p$ and $N_t/N_p$. It is therefore instructive to examine the compound ratio $N_t N_p / N_d^2$, for which these uncertainties largely cancel. 

\subsubsection{Compound Ratio $N_{t}N_{p}/N_{d}^{2}$}

\begin{figure}[h!] 
    \centering
    \includegraphics[width=.85\linewidth]{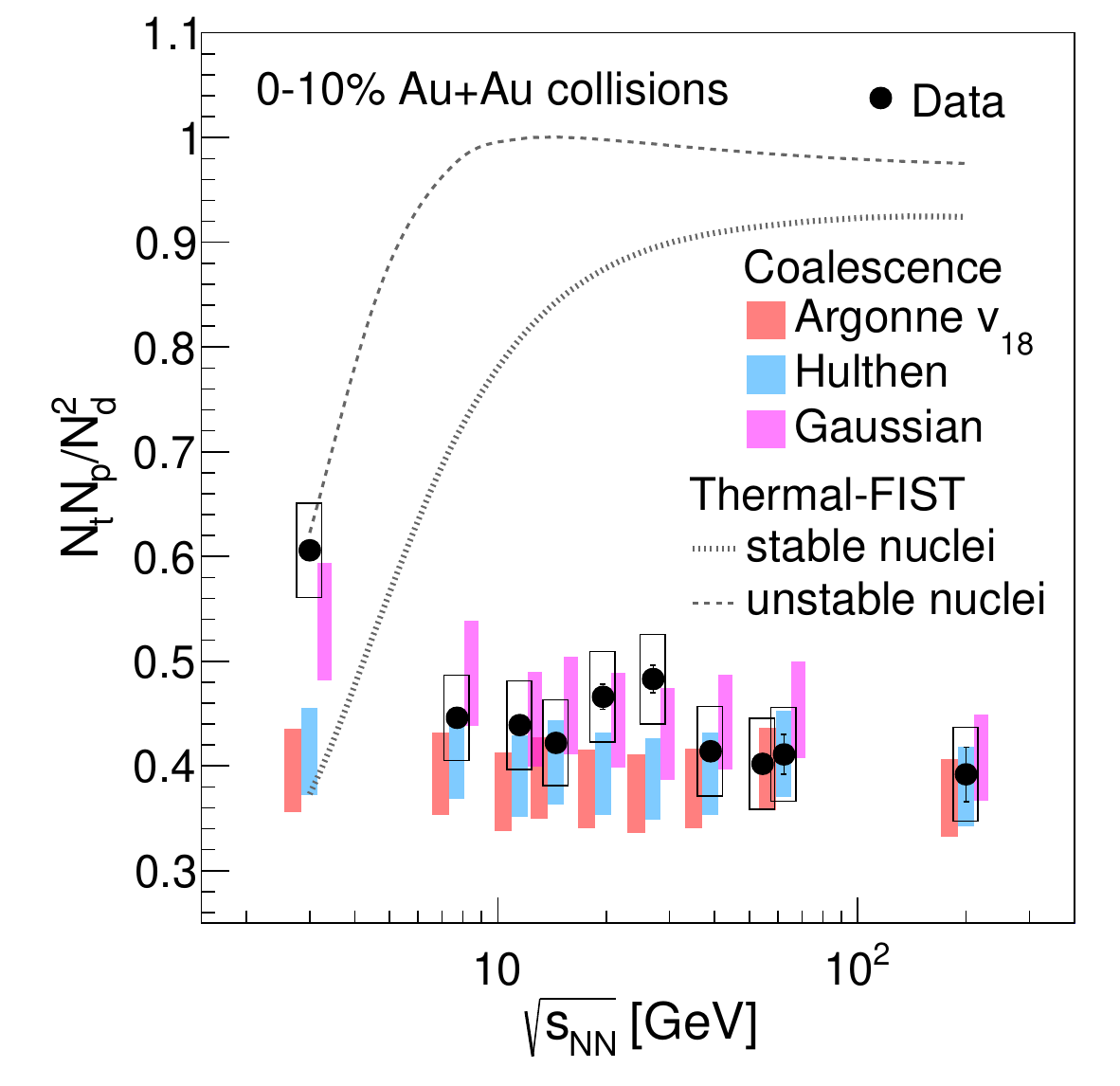} 
        \caption{The compound yield ratio $N_{t}N_{p}/N_{d}^{2}$ in $0-10\%$ Au+Au collisions as a function of $\sqrt{s_{\rm{NN}}}$. The data~\cite{STAR:2023uxk, STAR:2022hbp} are compared with coalescence (red bands) and Thermal-FIST~\cite{Vovchenko:2019pjl} (dashed lines) calculations.}
    \label{fig:tpd2_vs_snn}
\end{figure}

Figure~\ref{fig:tpd2_vs_snn} shows the compound ratio, $N_{t}N_{p}/N_{d}^{2}$ in $0-10\%$ Au+Au collisions. The data is compared with coalescence calculations using the three different deuteron wave functions, as well as Thermal-FIST calculations. The data at \snn=3 GeV is noticeably larger than those at \snn=7.7--200 GeV, which is roughly constant. Two thermal model calculations are shown. One using a particle list with only stable nuclei, and another including unstable nuclei and their decays. Both calculations overestimates the data from \snn=7.7--200 GeV by substantially, by approximately a factor of 2. The calculation including unstable nuclei is larger compared to that without, with the relative difference being larger at lower energies. This is mainly due to the stronger feed-down to to tritons at lower collision energies from unstable nuclei; the feed-down fraction is approximately $60\%$ for tritons at \snn=3 GeV according to the thermal model~\cite{Vovchenko:2020dmv}, enhancing the compound ratio strongly. 

The coalescence calculations give a reasonable description of the data. It is worth pointing out that the coalescence calculations Hulthen and Argonne $v_{18}$ both predict a almost constant dependence of $N_{t}N_{p}/N_{d}^{2}$ on energy, while the Gaussian predicts a slightly decreasing trend. The decreasing trend adopting the Gaussian wave function is also seen in calculations using transport model with coalescence afterburner with Gaussian ansatz as shown in Ref.~\cite{Zhao:2021dka}, and has been demonstrated to arise from the different spatial extent between the triton and deuteron wave functions. As the energy increases, the source radius increases, which also increases the relative suppression of the deuteron yield with respect to the triton yield arising from the wider wave function of the deuteron. However, this decreasing trend is not observed using the more realistic Hulthen and Argonne $v_{18}$ wave functions. This could be explained qualitatively: The Hulthen and Argonne $v_{18}$ wave functions feature a pronunced peak below 3 fm, whereas the Gaussian wave function is much more broad (see Ref.~\cite{Mahlein:2023fmx}). In other words, using the more realistic Hulthen and Argonne $v_{18}$ wave functions give rise to less suppression at small source sizes (e.g. \snn=3 GeV) compared to the Gaussian, driving the predicted $N_{d}$ up and $N_{t}N_{p}/N_{d}^{2}$ down. It should also be noted that the apparent disagreement at \snn=3 GeV could be partially due to the lack of consideration of feed-down components in our coalescence calculations.  

\subsubsection{Mean Transverse Momentum}
\begin{figure}[h!] 
    \centering
    \includegraphics[width=.85\linewidth]{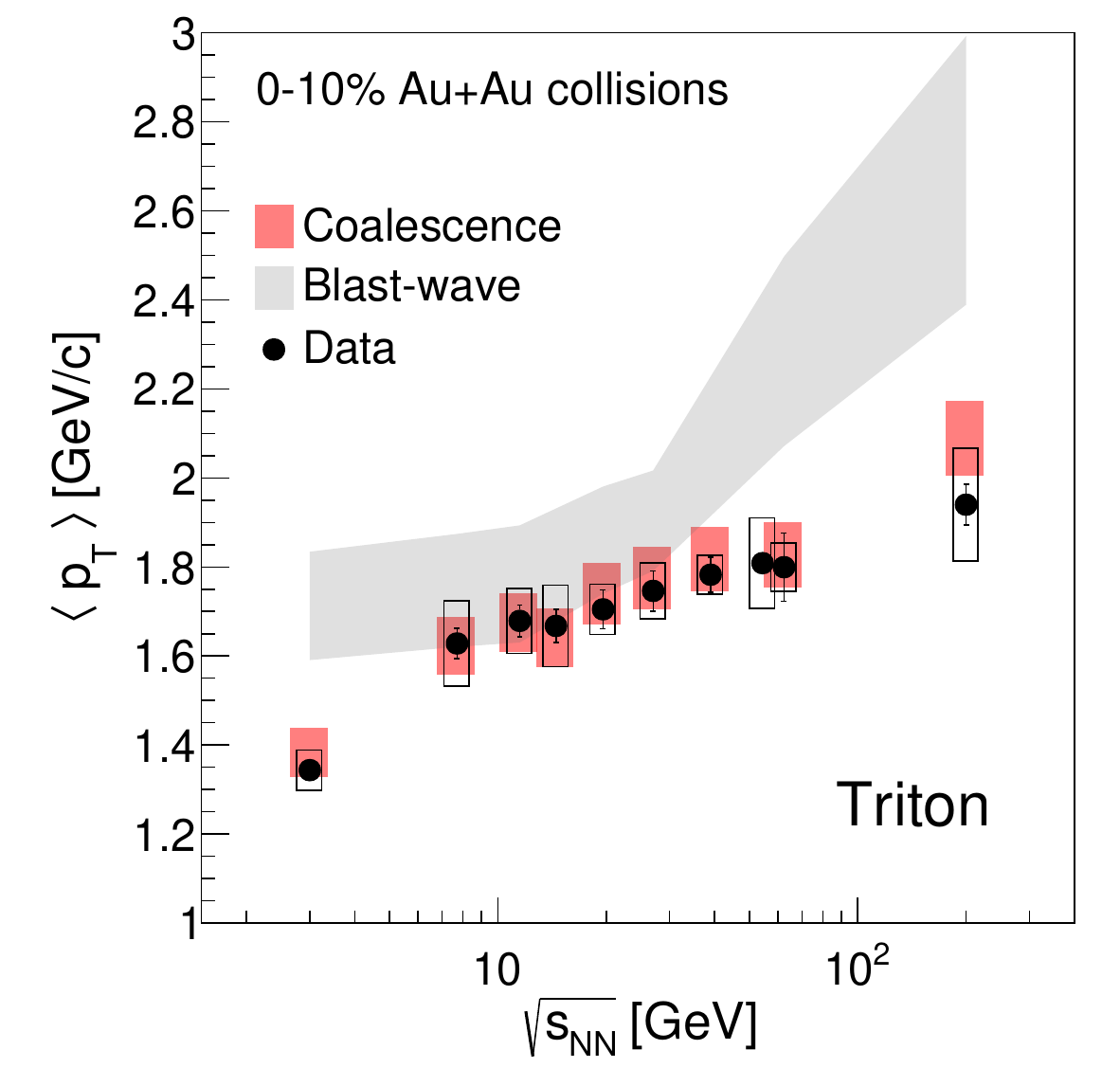} 
        \caption{Mean transverse momentum of tritons in $0-10\%$ Au+Au collisions as a function of $\sqrt{s_{\rm{NN}}}$. The data~\cite{STAR:2023uxk, STAR:2022hbp} are compared with coalescence (red bands) and blast-wave model (dashed lines) calculations where the kinetic freeze-out parameters are constrained by measured light hadron yields~\cite{STAR:2017sal}.}
    \label{fig:tmeanpt_vs_snn}
\end{figure}

Fig.~\ref{fig:tmeanpt_vs_snn} shows the mean transverse momentum of tritons as a function of collision energy. The data is compared with coaelscence calculations using the Argonne $v_{18}$ wave function. The results using the other two wave functions are similar and are not shown. The calculations describe the data very well. 

We also compare the data with a hydronamical-inspired blast-wave model. In a scenario where all particles, including nuclei, are subject to the same hydronimcal evolution, the $p_T$ spectra of all particles is expected to take the form of Eq.~\ref{eq:blast}. To generate blast-wave expectations for tritons, we utilize the published kinetic freeze-out parameters obtained by blast-wave fits to $\pi^\pm, K^\pm, p(\bar{p})$ for \snn=7.7--200 GeV~\cite{STAR:2017sal}. For \snn=3 GeV, since pion data is not yet available, we use the blast-wave fits to protons only. The blast-wave expectation tends to lie above the data points, and is most noticeable at \snn=3 GeV and 200 GeV. 

To summarize, by estimating the source size from existing deuteron and proton spectra at \snn = 3–200 GeV and employing coalescence calculations with realistic deuteron wave functions such as Argonne $v_{18}$ or Hulthen, we find that the same coalescence framework consistently describes the triton spectra across different collision centralities. Both the $p_{T}$-integrated yields and the mean transverse momentum, $\langle p_{T} \rangle$, are well reproduced. Comparisons with thermal and blast-wave models further demonstrate that tritons do not follow the same chemical or kinetic freeze-out surfaces as hadrons. We now turn our attention to the ${}^{3}_{\Lambda}\rm{H}$.


\subsection{${}^{3}_{\Lambda}\rm{H}$}
\subsubsection{Transverse Momentum Spectra}

\begin{figure}[h!] 
    \centering
    \includegraphics[width=.99\linewidth]{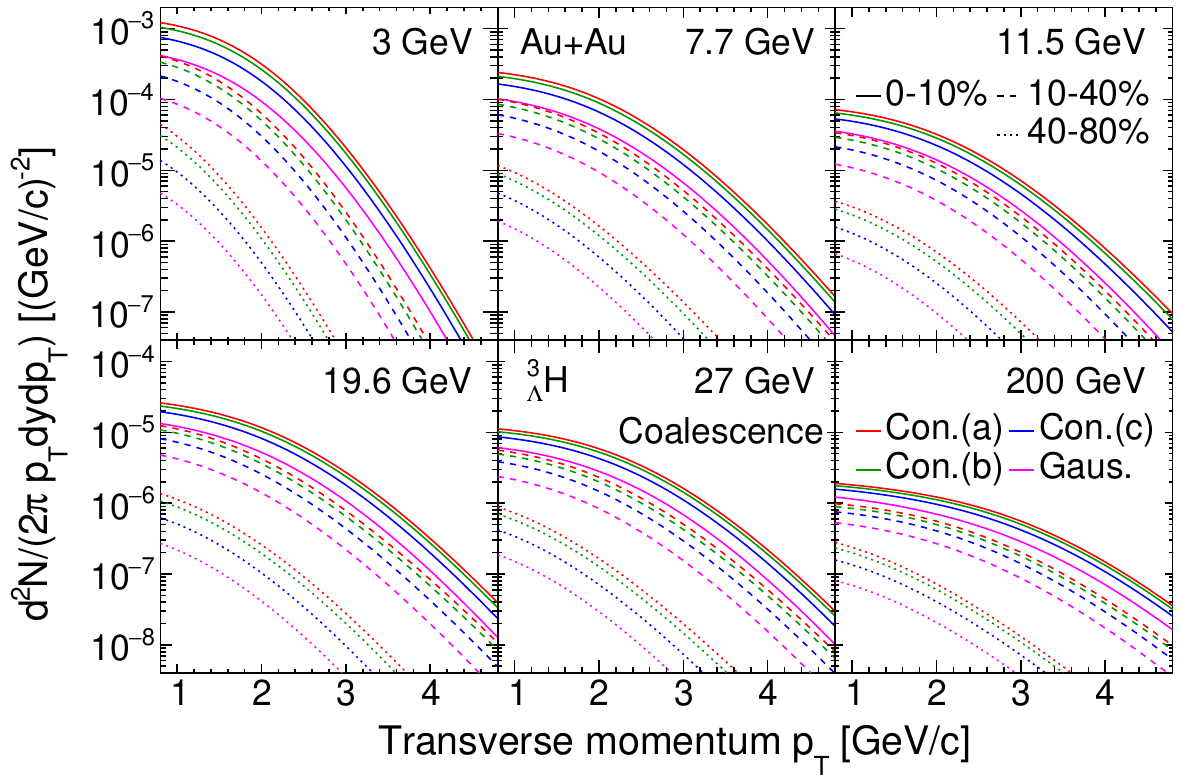} 
    \caption{${}^{3}_{\Lambda}\rm{H}$ transverse momentum spectra at \snn=3, 7.7, 11.5, 19.6, 27, and 200 GeV in $0-10\%$, $10-40\%$, and $40-80\%$ predicted from coalescence calculations using different ${}^{3}_{\Lambda}\rm{H}$ wave functions. The uncertainty on the calculations are not shown.}
    \label{fig:hypertriton_spectra_coal}
\end{figure}

Figure~\ref{fig:hypertriton_spectra_coal} presents the coalescence calculations for the $p_{T}$ spectra of ${}^{3}_{\Lambda}\rm{H}$ at \snn = 3, 7.7, 11.5, 19.6, 27, and 200 GeV for three centrality classes ($0$–$10\%$, $10$–$40\%$, and $40$–$80\%$). The source size is determined using the Argonne $v_{18}$ deuteron wave function. Different colored lines correspond to calculations with various ${}^{3}_{\Lambda}\rm{H}$ wave functions: the red, green, and blue curves represent the Congleton wave function with three parameterizations (see Tab.~\ref{tab:congparams}), while the magenta curve corresponds to a Gaussian wave function. For clarity, the uncertainties associated with source size variation are omitted; their relative magnitude is comparable to those of the triton spectra discussed in the previous section. 

Several qualitative observations can be made from the predicted spectra. First, the ${}^{3}_{\Lambda}\rm{H}$ yield exhibits a strong dependence on the choice of wave function. The Congleton (a) wave function consistently predicts the largest yield across all collision energies and centralities, whereas the Gaussian wave function yields the smallest. This trend can be understood by considering, for example, the $0$–$10\%$ centrality class at \snn = 200 GeV. The source radius in these collisions is estimated to be about 4.5 fm (Fig.~\ref{fig:rinv_vs_snn}). Since the coalescence probability is proportional to the overlap between the source and the wave function, wave functions with larger probability density at $r_{d\Lambda}<4.5$ fm will yield higher production rates. Among the considered cases, the Congleton (a) wave function provides the largest overlap with the source, while the Gaussian provides the smallest.  

Second, the differences between wave functions are most pronounced in peripheral collisions and at lower energies. This behavior is also intuitive: both decreasing beam energy and increasing collision centrality lead to smaller source radii. As shown in Fig.~\ref{fig:congleton_wave_function}, the discrepancies between the wave functions grow as $r_{d\Lambda}$ decreases below 4.5 fm. Consequently, the relative variation in overlap probability becomes larger, leading to stronger differences in predicted yields at low energies and in peripheral events. In other words, smaller collision systems provide particularly favorable conditions for probing the structure of the ${}^{3}_{\Lambda}\rm{H}$ wave function.  

\begin{figure}[h!] 
    \centering
    \includegraphics[width=.85\linewidth]{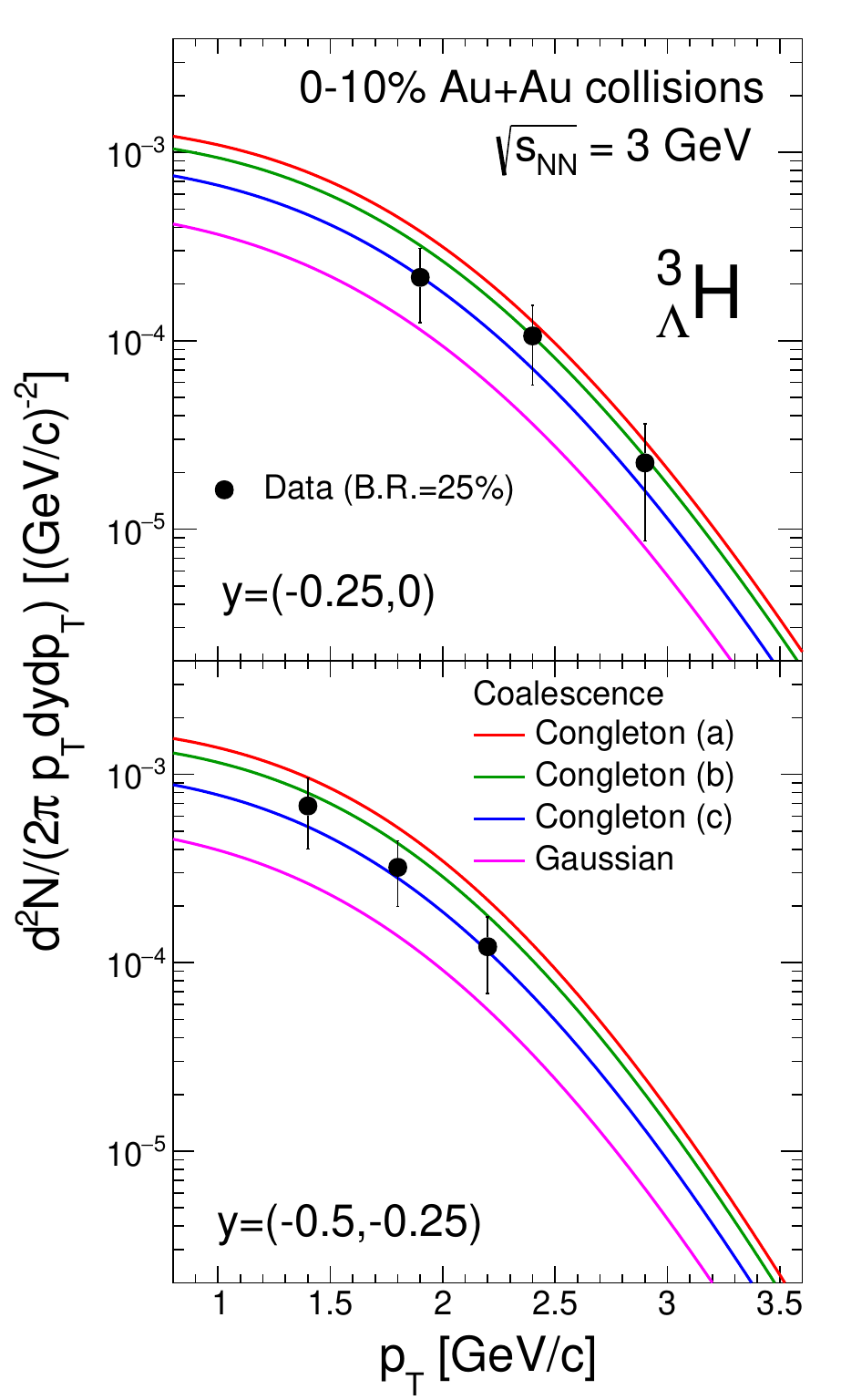} 
    \caption{${}^{3}_{\Lambda}\rm{H}$ transverse momentum spectra in $0-10\%$ \snn=3 GeV collisions. The data~\cite{STAR:2021orx}, in two rapidity regions $(-0.25,0)$ (upper panel) and $(-0.5,-0.25)$ (lower panel) are compared with coalescence calculations using different ${}^{3}_{\Lambda}\rm{H}$ wave functions are shown. The uncertainty on the calculations are not shown.}
    \label{fig:h3l_spectra_3gev_y4}
\end{figure}

The above predictions are for mid-rapidity. The STAR Collaboration has reported ${}^{3}_{\Lambda}\mathrm{H}$ yields in $0$--$10\%$ Au+Au collisions at $\sqrt{s_{\mathrm{NN}}}=3~\mathrm{GeV}$ in two rapidity intervals, $(-0.25,0)$ and $(-0.5,-0.25)$~\cite{STAR:2021orx}. A branching ratio of $25\%$ for the decay ${}^{3}_{\Lambda}\mathrm{H}\rightarrow {}^{3}\mathrm{He}+\pi^{-}$ is assumed in the data. By repeating our procedure in these two rapidity bins, i.e. using the deuteron, proton, and $\Lambda$ spectra in these rapidity bins, we obtain predictions for the ${}^{3}_{\Lambda}\mathrm{H}$ yield, which are compared with the data in Fig.~\ref{fig:h3l_spectra_3gev_y4}. The uncertainty associated with the source size, which is of the order $10\%$, is not shown. 

Overall, we find that the calculations based on the Congleton (b) and (c) wave functions are in good agreement with the data, while the Congleton (a) results tend to lie slightly above the measurements and the Gaussian wave function results tend to lie slightly below them. These comparisons may indicate that the Gaussian wave function is too broad, corresponding to a suppressed probability at $r_{d\Lambda}<3~\mathrm{fm}$, where 3 fm is the estimated source radius in $0$--$10\%$ Au+Au collisions at $\sqrt{s_{\mathrm{NN}}}=3~\mathrm{GeV}$. However, more precise data will be required to establish this conclusively. 

We have also repeated the calculations using a set of more compact wave functions with $\langle r_{d\Lambda} \rangle=9.1\,\mathrm{fm}$ (see Appendix~\ref{sec:appendix_2} for details), obtained by modifying the parameters $b_{\Lambda}$ and $\alpha_{\Lambda}$ for the Gaussian and Congleton wave functions, respectively. This results in an increase of the predicted yields by approximately $15$--$40\%$ for all wave functions. The enhancement arises because reducing $\langle r_{d\Lambda} \rangle$ shifts the wave-function probability toward smaller separations, thereby increasing its overlap with the nucleon emission source at $\sqrt{s_{\mathrm{NN}}}=3~\mathrm{GeV}$.

\subsubsection{Centrality Dependence}

\begin{figure}[h!] 
    \centering
    \includegraphics[width=.85\linewidth]{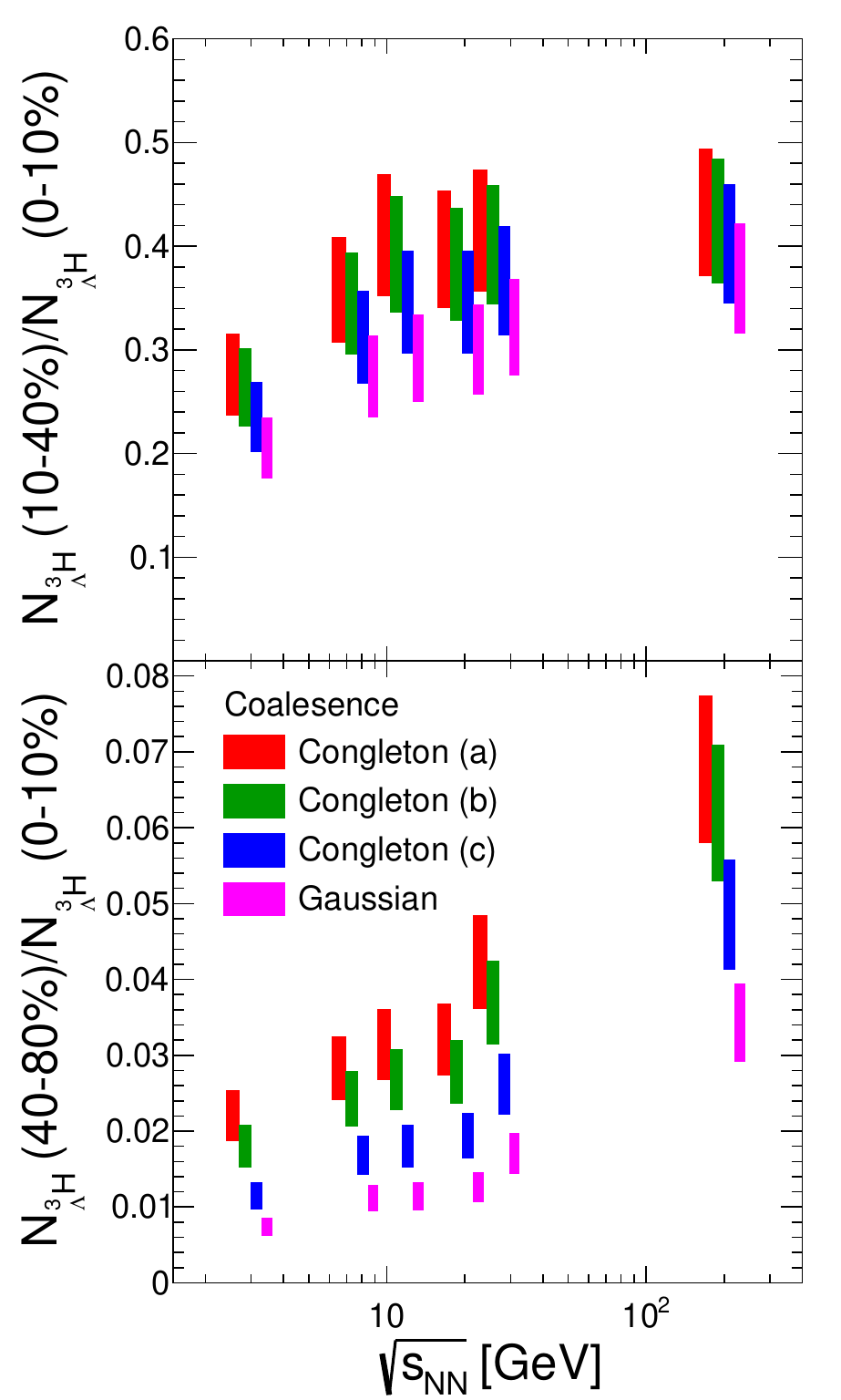} 
        \caption{Ratio of ${}^{3}_{\Lambda}\rm{H}$ yields in $10$-$40\%$ centrality (upper panel) and $40$-$80\%$ centrality (lower panel) relative to $0-10\%$. Different colors represent coalescence calculations using different ${}^{3}_{\Lambda}\rm{H}$ wave functinos.}
    \label{fig:rcp}
\end{figure}

It is instructive to examine the centrality dependence of ${}^{3}_{\Lambda}\rm{H}$ production. The HADES collaboration has observed a remarkably steep dependence in \snn=2.55 GeV Ag+Ag collisions, significantly stronger than that seen for protons and $\Lambda$ hyperons~\cite{hades}. In other words, the increase in ${}^{3}_{\Lambda}\rm{H}$ yield from peripheral to central collisions exceeds that of its constituents. Figure~\ref{fig:rcp} shows the yield ratios in the $10$–$40\%$ and $40$–$80\%$ centrality classes relative to $0$–$10\%$. In both cases, the ratios rise with energy from \snn=3 to 7.7 GeV, remain roughly constant, and then increase again from 27 to 200 GeV. This trend reflects the collision-energy dependence of the source radius shown in Fig.~\ref{fig:rinv_vs_snn}: smaller ratios are observed for smaller source sizes. Moreover, the ratios are sensitive to the choice of the hypernuclear wavefunction, especially at low collision energies. This sensitivity underscores the potential of yield measurements at low and intermediate \snn\ to serve as a precision tool for probing the structure of hypernuclei. 

\subsubsection{$p_{T}$-Integrated Hypernuclei-to-Hyperon Yield Ratios}

\begin{figure*}[t] 
    \centering
    \includegraphics[width=.7\linewidth]{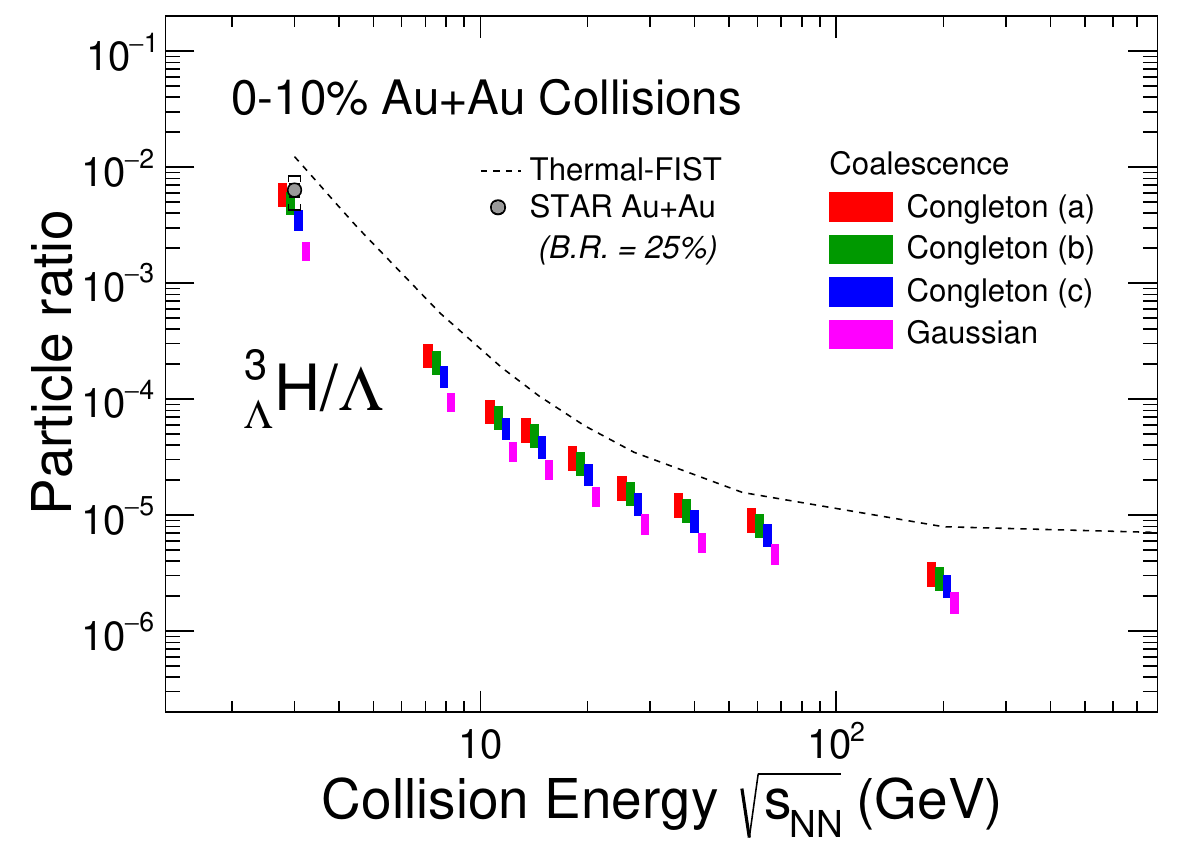} 
        \caption{Yield ratios $N_{{}^{3}_{\Lambda}\rm{H}}/N_{\Lambda}$ in $0-10\%$ Au+Au collisions as a function of $\sqrt{s_{\rm{NN}}}$. The data~\cite{STAR:2021orx, STAR:2024znc} are compared with coalescence (colored bands) and Thermal-FIST~\cite{Vovchenko:2019pjl} (dashed lines) calculations. }
    \label{fig:h3lld_vs_snn}
\end{figure*}

Figure~\ref{fig:h3lld_vs_snn} presents the yield ratio 
$N_{{}^{3}_{\Lambda}\mathrm{H}}/N_{\Lambda}$ in $0$--$10\%$ Au+Au collisions as a function of $\sqrt{s_{\mathrm{NN}}}$, compared with predictions from coalescence and thermal models. 
The thermal results are obtained with the Thermal-FIST package~\cite{Vovchenko:2019pjl}, using freeze-out parameters $(T,\mu_B)$ from Ref.~\cite{Vovchenko:2015idt}. 
Similar to what has been observed for tritons, the coalescence predictions for $N_{{}^{3}_{\Lambda}\mathrm{H}}/N_{\Lambda}$ lie systematically below the thermal model results. 
The relative difference between the two approaches remains approximately constant across energies. 
Consistent with the behavior of the $p_{T}$ spectra, the predicted ratio depends on the choice of the wave function, with the Congleton (a) wave function giving the largest ratio and the Gaussian wave function the lowest. 
At $\sqrt{s_{\mathrm{NN}}}=3$~GeV, the Gaussian prediction underestimates the data, whereas the three Congleton wave functions provide better descriptions.

\subsubsection{Strangeness Population Factor $S_{3}=(N_{{}^{3}_{\Lambda}\rm{H}}/N_{\Lambda})/(N_{{}^{3}_{}\rm{H}e}/N_{p})$ }

The ratio $S_{3}=(N_{{}^{3}_{\Lambda}\rm{H}}/N_{\Lambda})/(N_{{}^{3}_{}\rm{He}}/N_{p})$ is a useful observable because it cancels out the effects of differences in nucleon and hyperon densities, and can directly quantify the relative production probabilities of nuclei (${}^{3}\rm{He}$) and hypernuclei (${}^{3}_{\Lambda}\rm{H}$). It also offers the advantage that certain theoretical and experimental uncertainties may partially cancel between the numerator and denominator. 

\begin{figure}[h!] 
    \centering
    \includegraphics[width=.85\linewidth]{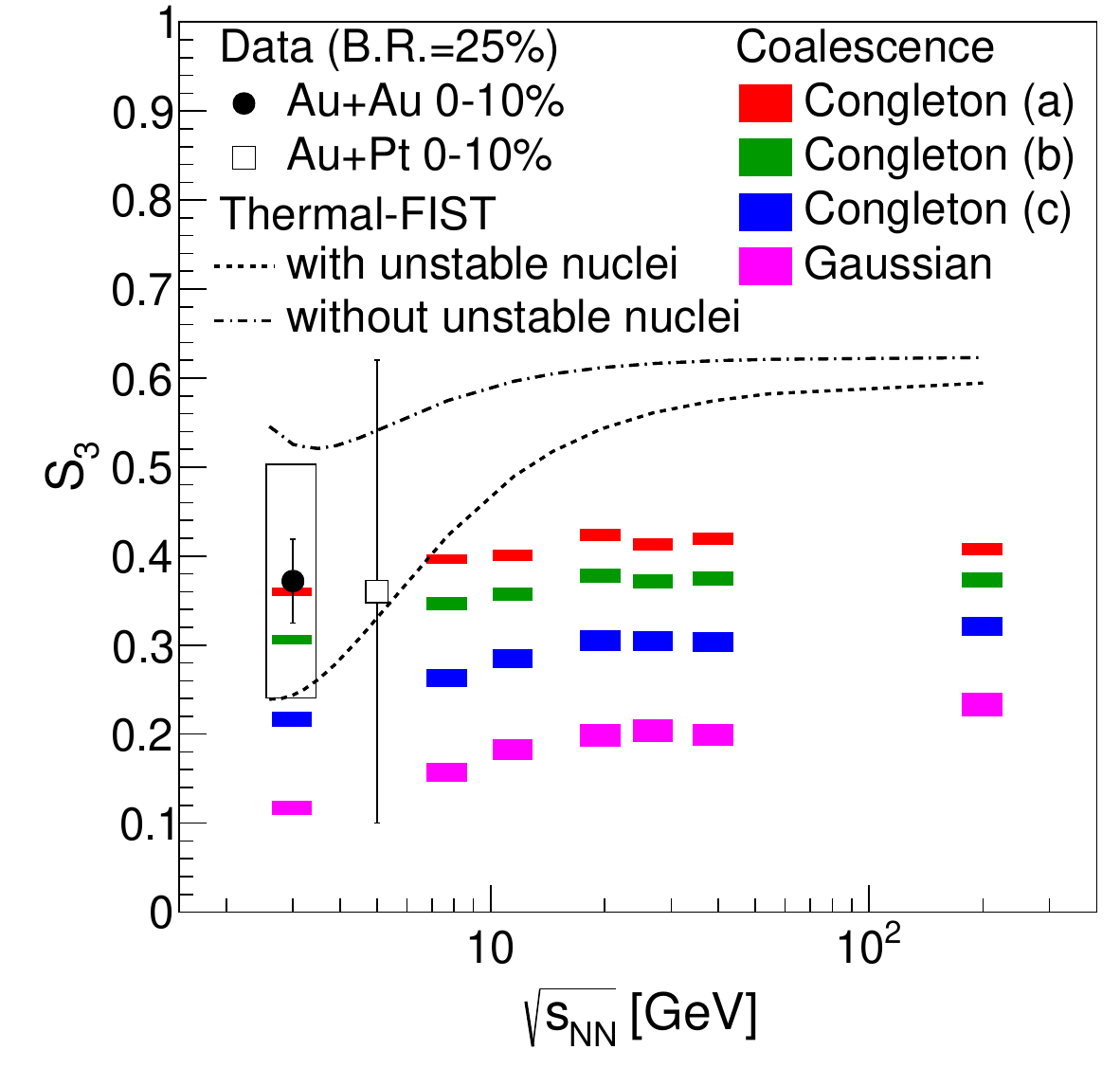} 
        \caption{$S_{3}$ in heavy-ion collisions as a function of $\sqrt{s_{\rm{NN}}}$. The data~\cite{STAR:2021orx, STAR:2023uxk, STAR:2024znc, E864:2002xhb} are compared with coalescence calculations using different ${}^{3}_{\Lambda}\rm{H}$ wave functions (colored bands) and Thermal-FIST calculations~\cite{Vovchenko:2019pjl} (dashed lines) including and not including unstable nuclei.}
    \label{fig:s3_vs_snn}
\end{figure}

Figure~\ref{fig:s3_vs_snn} shows $S_{3}$ as a function of collision energy in $0$–$10\%$ Au+Au collisions. For the coalescence calculations, we note that the uncertainty originating from the source size is significantly reduced, allowing a more transparent interpretation. The calculations predict a mildly increasing trend in $S_{3}$ from $\sqrt{s_{\rm{NN}}}=3$ to $11.5$~GeV, after which it remains approximately constant up to 200~GeV. Furthermore, the difference between results obtained with different wave functions becomes more pronounced at lower energies; collisions with smaller source radii are more sensitive to wave-function effects. This observation underscores the potential of low-energy collisions as a probe of the hypernuclear wave function.

For the thermal model calculations, we consider two scenarios: one including only stable nuclei, and another also accounting for the decay of unstable nuclei. A large difference is observed between the two, especially at energies below $\sqrt{s_{\rm{NN}}}=7.7$~GeV. This discrepancy does not arise from any decay channels feeding into ${}^{3}_{\Lambda}\rm{H}$, since no such channels are known. Instead, it is attributed to the decay of unstable $A=4$ nuclei into ${}^{3}\rm{He}$~\cite{Vovchenko:2020dmv}. However, these contributions are likely overestimated: the E864 Collaboration has demonstrated that in $\sqrt{s_{\rm{NN}}}=4.8$~GeV Au+Pt collisions, the yields of unstable nuclei such as ${}^{4}\rm{H}$ and ${}^{4}\rm{Li}$ are substantially lower than thermal model predictions~\cite{Armstrong:2001mr}. A more careful treatment of such feed-down effects may need to be incorporated into the coalescence framework in future studies for a precise interpretation of $S_{3}$ at lower collision energies.

Experimental data are still scarce. In addition to the $0$–$10\%$ Au+Au measurements at $\sqrt{s_{\rm{NN}}}=3$~GeV~\cite{STAR:2021orx, STAR:2024znc, STAR:2023uxk}, we compare with results from $0$–$10\%$ Au+Pt collisions measured by the AGS E864 experiment~\cite{E864:2002xhb}. Unfortunately, the experimental uncertainties remain substantial, precluding a definitive conclusion. Future measurements from the STAR Beam Energy Scan program will provide crucial input for further constraining these studies.

We have mentioned that low collision energies provide a promising avenue for studying the ${}^{3}_{\Lambda}\mathrm{H}$. At the same time, small systems at high collision energies also offer distinct advantages. First, they are essentially free from unstable-nuclei feed-down owing to the strong suppression factor. Second, the source size has been accurately measured by the ALICE collaboration as a function of $m_{T}$~\cite{ALICE:2025wuy} and shown to scale with $m_{T}$ for different baryons, indicating a common baryon source~\cite{ALICE:2020ibs}.

In view of the forthcoming STAR data from Ru+Ru and Zr+Zr collisions at $\sqrt{s_{\mathrm{NN}}}=200$~GeV, we calculate the $S_{3}$ ratio as a function of charged-particle multiplicity. The characteristic $\langle m_{T} \rangle$ of protons in Ru+Ru and Zr+Zr collisions at this energy is approximately $1.3~\mathrm{GeV}/c^{2}$. Since the source size in high-energy measurements has already been quantified via femtoscopic studies by ALICE, for this prediction, we directly use their parameterized data rather using our method. Specifically, for $m_{T}=1.3$--$1.4~\mathrm{GeV}/c^{2}$, we use $R_{\mathrm{inv}} = 0.428 \, \langle dN_{\mathrm{ch}}/d\eta \rangle^{1/3}$ from Ref.~\cite{ALICE:2025wuy}. 

\begin{figure}[h!] 
    \centering
    \includegraphics[width=.85\linewidth]{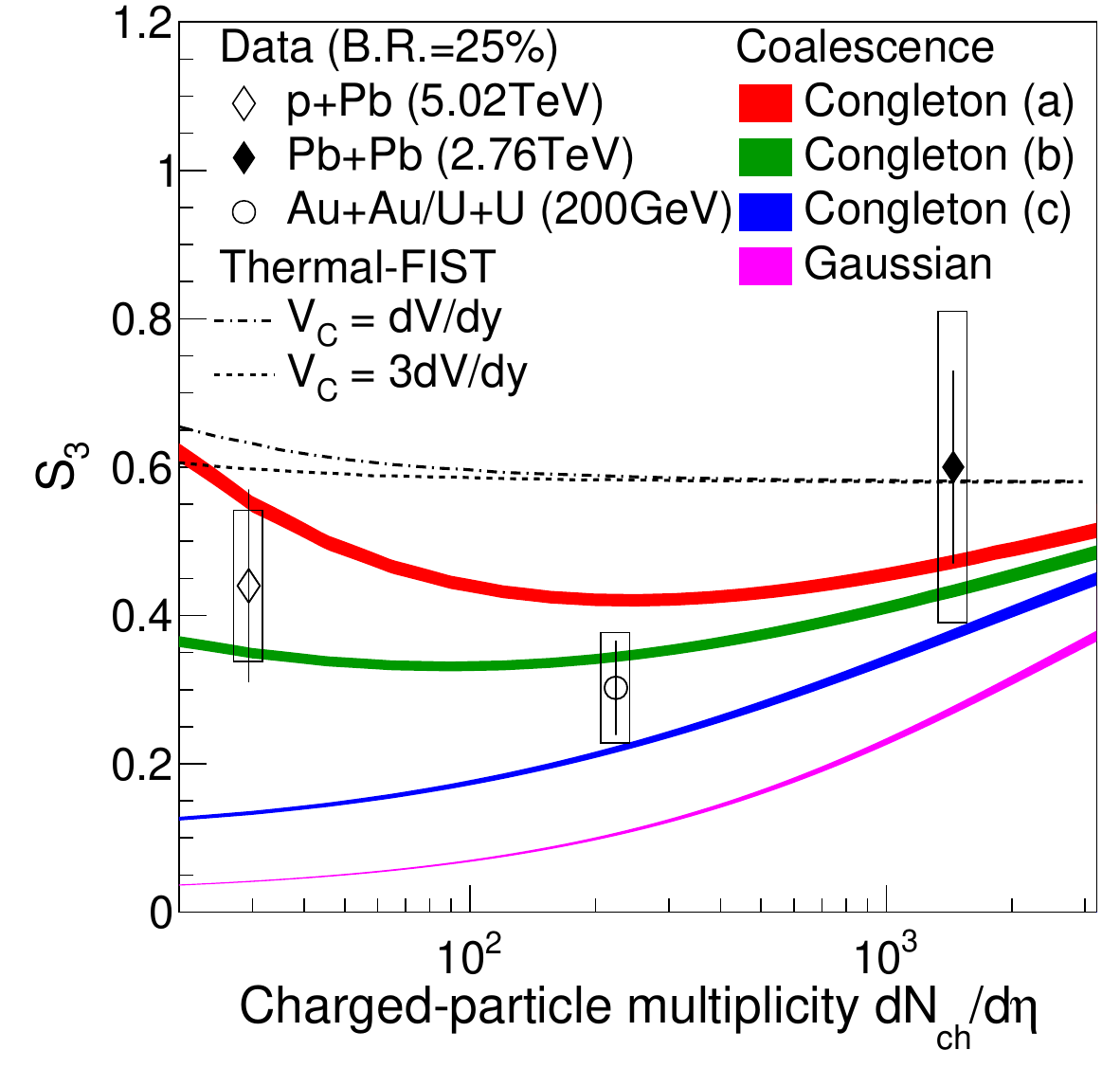} 
        \caption{$S_{3}$ in heavy-ion collisions as a function of charged-particle multiplicity $dN_{\rm{ch}}/d\eta$. The data~\cite{ALargeIonColliderExperiment:2021puh, STAR:2023fbc, ALICE:2015oer} are compared with coalescence calculations using different ${}^{3}_{\Lambda}\rm{H}$ wave functions (colored bands) and Thermal-FIST calculations~\cite{Vovchenko:2019pjl} (dashed lines).}
    \label{fig:s3_vs_nch}
\end{figure}

Figure~\ref{fig:s3_vs_nch} presents the results of these calculations using different wave functions. At high multiplicities the calculations from different wave functions converge, whereas at low multiplicities they show increasing deviations. This again highlights the potential of small systems for probing the ${}^{3}_{\Lambda}\mathrm{H}$ wave function. The Ru+Ru and Zr+Zr systems at $\sqrt{s_{\mathrm{NN}}}=200$~GeV span the multiplicity range $20 \lesssim \langle dN_{\mathrm{ch}}/d\eta \rangle \lesssim 400$, providing a valuable opportunity for such studies. 

For comparison, we also include thermal-model predictions within the canonical ensemble using correlation volumes $V_{c}=dV/dy$ and $V_{c}=3dV/dy$. Both variants yield a constant value of $S_{3}\approx 0.6$ across the multiplicity range $20$--$2000$. 

For reference, the measurements from $\sqrt{s_{\mathrm{NN}}}=2.76$~TeV 0--10\% Pb+Pb collisions~\cite{ALICE:2015oer}, $\sqrt{s_{\mathrm{NN}}}=200$~GeV (193~GeV) Au+Au (U+U) minimum-bias collisions~\cite{STAR:2023fbc}, and $p$+Pb collisions at $\sqrt{s_{\mathrm{NN}}}=5.02$~TeV~\cite{ALargeIonColliderExperiment:2021puh} are shown. Notably, both the $p$+Pb and Au+Au (U+U) data disfavor the Gaussian wave function, consistent with the observation at $\sqrt{s_{\mathrm{NN}}}=3$~GeV. Instead, the data show a preference for the Congleton (b) wave function. However, the current uncertainties remain substantial and prevent a definitive conclusion.

We have also repeated the calculations using a set of more compact wave functions with $\langle r_{d\Lambda} \rangle=9.1\,\mathrm{fm}$ (see Appendix~\ref{sec:appendix_2} for details), following the same procedure as that adopted previously for the predictions at $\sqrt{s_{\mathrm{NN}}}=3~\mathrm{GeV}$. This leads to a slight increase in the predicted $S_{3}$ values for all wave functions. Nevertheless, the results using the Gaussian wave function remains the lowest among the four and continues to lie below the $p$+Pb and Au+Au (U+U) data.

\subsubsection{Mean Transverse Momentum}
\begin{figure}[h!] 
    \centering
    \includegraphics[width=.85\linewidth]{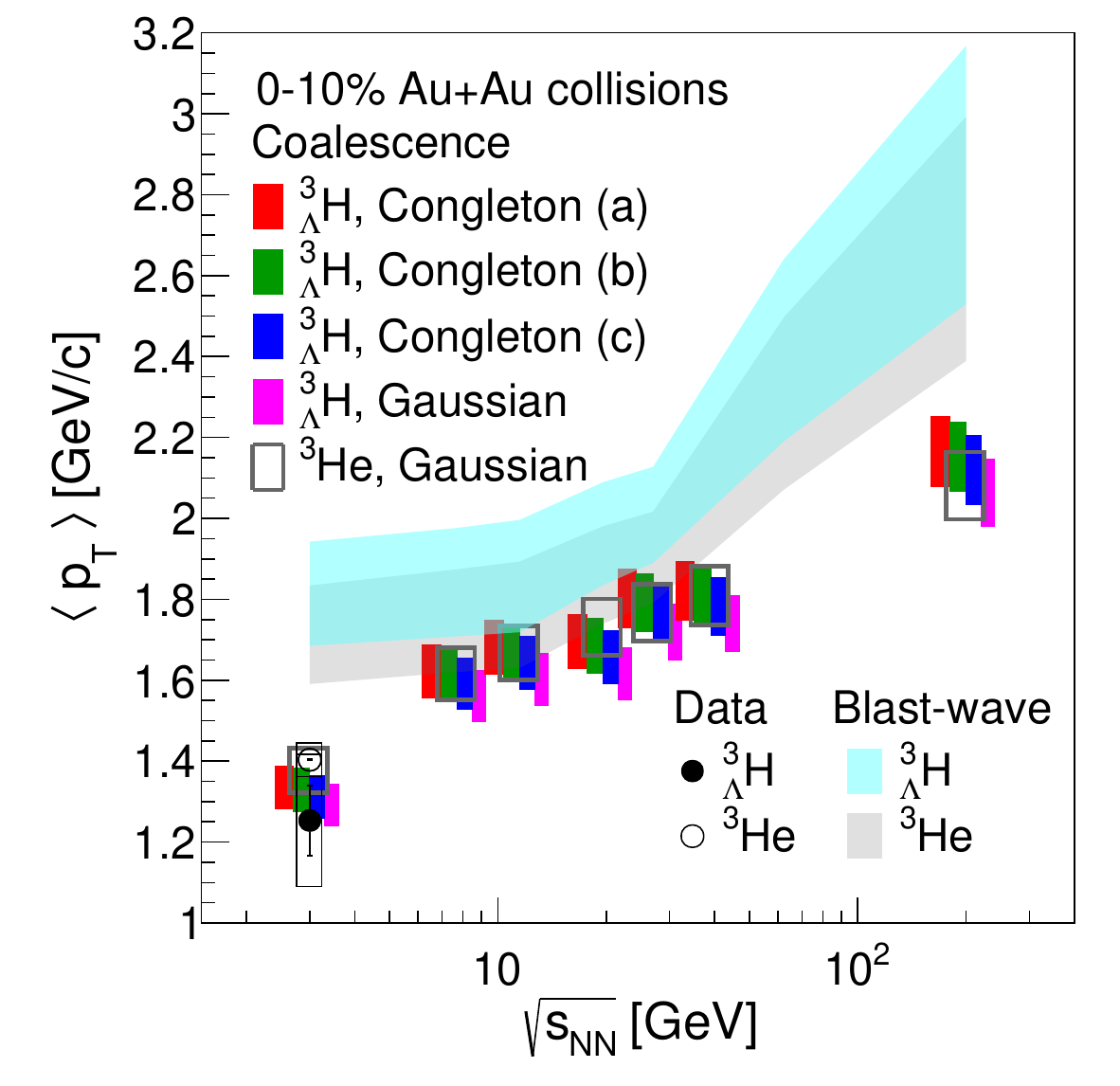} 
        \caption{Mean transverse momentum of ${}^{3}_{\Lambda}\rm{H}$ in $0-10\%$ Au+Au collisions as a function of $\sqrt{s_{\rm{NN}}}$. The data~\cite{STAR:2021orx} are compared with coalescence (colored bars) and blast-wave model (pink band) calculations where the kinetic freeze-out parameters are constrained by measured light hadron yields~\cite{STAR:2017sal}. The corresponding data~\cite{STAR:2023uxk} and calculations for ${}^{3}\rm{He}$ are also shown for comparison. }
    \label{fig:h3lmeanpt_vs_snn}
\end{figure}

Finally, we examine the mean transverse momentum of ${}^{3}_{\Lambda}\rm{H}$ as a function of collision energy. Figure~\ref{fig:h3lmeanpt_vs_snn} shows the $\langle p_{T} \rangle$ of ${}^{3}_{\Lambda}\rm{H}$ in $0$--$10\%$ Au+Au collisions at various collision energies. The coalescence calculations, shown as the colored bars, exhibit little dependence on the choice of wave function, in contrast to the yield results. For comparison, the corresponding coalescence calculations for ${}^{3}\rm{He}$ are shown as gray rectangles. The mean transverse momenta of ${}^{3}\rm{He}$ and ${}^{3}_{\Lambda}\rm{H}$ are found to be very similar across all energies. This similarity arises because the $\langle p_{T} \rangle$ of nuclei and hypernuclei is largely governed by the momentum spectra of their constituent nucleons and hyperons, which are themselves quite similar. For instance, at $\sqrt{s_{\rm{NN}}}=200$~GeV the mean transverse momentum of the $\Lambda$ exceeds that of the proton by about $10\%$, whereas at $\sqrt{s_{\rm{NN}}}=3$~GeV the opposite trend is observed~\cite{STAR:2024znc}. At intermediate energies, the differences remain below $10\%$. A small suppression of the ${}^{3}_{\Lambda}\rm{H}$ $\langle p_{T} \rangle$ is expected due to its broader wave function compared to ${}^{3}\rm{He}$, as also noted in Ref.~\cite{Liu:2024ygk}. However, this effect may be difficult to resolve due to experimental uncertainties. 

We also present blast-wave predictions for ${}^{3}_{\Lambda}\rm{H}$ and ${}^{3}\rm{He}$, shown as cyan and gray bands, respectively. The blast-wave parameters are obtained from fits to $\pi^{\pm}$, $K(\bar{K})$, and $p(\bar{p})$ spectra. The blast-wave predictions for ${}^{3}_{\Lambda}\rm{H}$ are slightly higher than those for ${}^{3}\rm{He}$, reflecting the slightly larger mass of ${}^{3}_{\Lambda}\rm{H}$. Notably, the blast-wave predictions for ${}^{3}_{\Lambda}\rm{H}$ lie systematically above the coalescence results, independent of the wave function choice. 

Finally, we compare our calculations with the ${}^{3}_{\Lambda}\rm{H}$ and ${}^{3}\rm{He}$ data at $\sqrt{s_{\rm{NN}}}=3$~GeV~\cite{STAR:2021orx, STAR:2023uxk}. The measurements are consistent with the coalescence calculations for both ${}^{3}_{\Lambda}\rm{H}$ and ${}^{3}\rm{He}$, while being underestimated by the blast-wave model. This observation echoes the conclusion drawn earlier from the triton results: $A=3$ hypernuclei do not share the same kinetic freeze-out surface as hadrons, and the coalescence model provides a natural explanation for this behavior.

\section{Summary and Outlook}
\label{sec:summary}

In this work, we have developed and applied a data-guided coalescence framework for the production of $A=3$ nuclei and hypernuclei. Within this approach, the source radius is extracted from experimental proton and deuteron yields, in conjunction with coalescence calculations. The resulting source size in Au+Au collisions at $\sqrt{s_{\rm{NN}}}=3$--200~GeV and Pb+Pb collisions at $\sqrt{s_{\rm{NN}}}=5.02$~TeV exhibits an approximate scaling with charged-particle multiplicity.

Using the extracted source size, the coalescence model was then employed to predict triton, ${}^{3}\mathrm{He}$, and ${}^{3}_{\Lambda}\mathrm{H}$ observables in Au+Au collisions across the same energy range. For tritons, calculations with a Gaussian wave function reproduce the measured transverse-momentum spectra, yield ratios, and mean transverse momenta. In contrast, thermal model calculations and blast-wave predictions are less successful in describing these observables.

For the hypertriton (${}^{3}_{\Lambda}\mathrm{H}$), four different parameterizations were tested: three based on the Congleton wave function and one Gaussian form. Although all four wave functions exhibit a similar RMS deuteron--$\Lambda$ separation of about $10.8\,\mathrm{fm}$, the resulting model predictions were found to be highly sensitive to the specific choice of wave function. The sensitivity is most pronounced at low collision energies and in peripheral collisions, where the source volume is small. Among the tested forms, the Congleton wave function with parameterization (b), which treats the hypertriton as a two-body system, provides a good description of ${}^{3}_{\Lambda}\mathrm{H}$ spectra at $\sqrt{s_{\rm{NN}}}=3$~GeV, while the Gaussian form slightly underestimates the data. The strangeness population factor $S_{3}$ also exhibits dependence on the hypertriton wave function, particularly in low-multiplicity environments. Comparisons with $p$+Pb data at the LHC and with Au+Au and U+U data at the top RHIC energy indicate that Gaussian-based coalescence calculations tend to underestimate ${}^{3}_{\Lambda}\mathrm{H}$ production. More precise data are needed to draw a firm conclusion. These findings highlight the strong sensitivity of hypernuclear observables to the assumed short-distance structure of the wave function, and they underline the unique potential of low-energy heavy-ion collisions to probe hypernuclear structure.

Looking ahead, several natural extensions of this work could provide deeper theoretical insight. A fully dynamical transport model can replace the simplified source distribution parameterization. A more accurate description of the source, including spatial and temporal correlations, collective flow, and species-dependent freeze-out dynamics, will help disentange the roles of wave function structure and medium effects in hypernuclear formation. Furthermore, the methodology can be extended to study the production of heavier hypernuclei such as ${}^{4}_{\Lambda}\mathrm{H}$, ${}^{4}_{\Lambda}\mathrm{He}$, and ${}^{5}_{\Lambda}\mathrm{He}$. These systems, with richer structure and binding dynamics, may provide additional constraints on the multi-body hyperon--nucleon interactions at short distances. 

\section*{Acknowledgements}
The authors are grateful to Jiaxing Zhao, Maximiliano Puccio, and Maximilian Mahlein for insightful discussions, and to Kfir Blum for generously sharing his coalescence code.

\clearpage
\appendix

\section{Radial probability distributions of the ${}^{3}_{\Lambda}\rm{H}$ wave functions}
\begin{figure}[h!] 
    \centering
    \includegraphics[width=.9\linewidth]{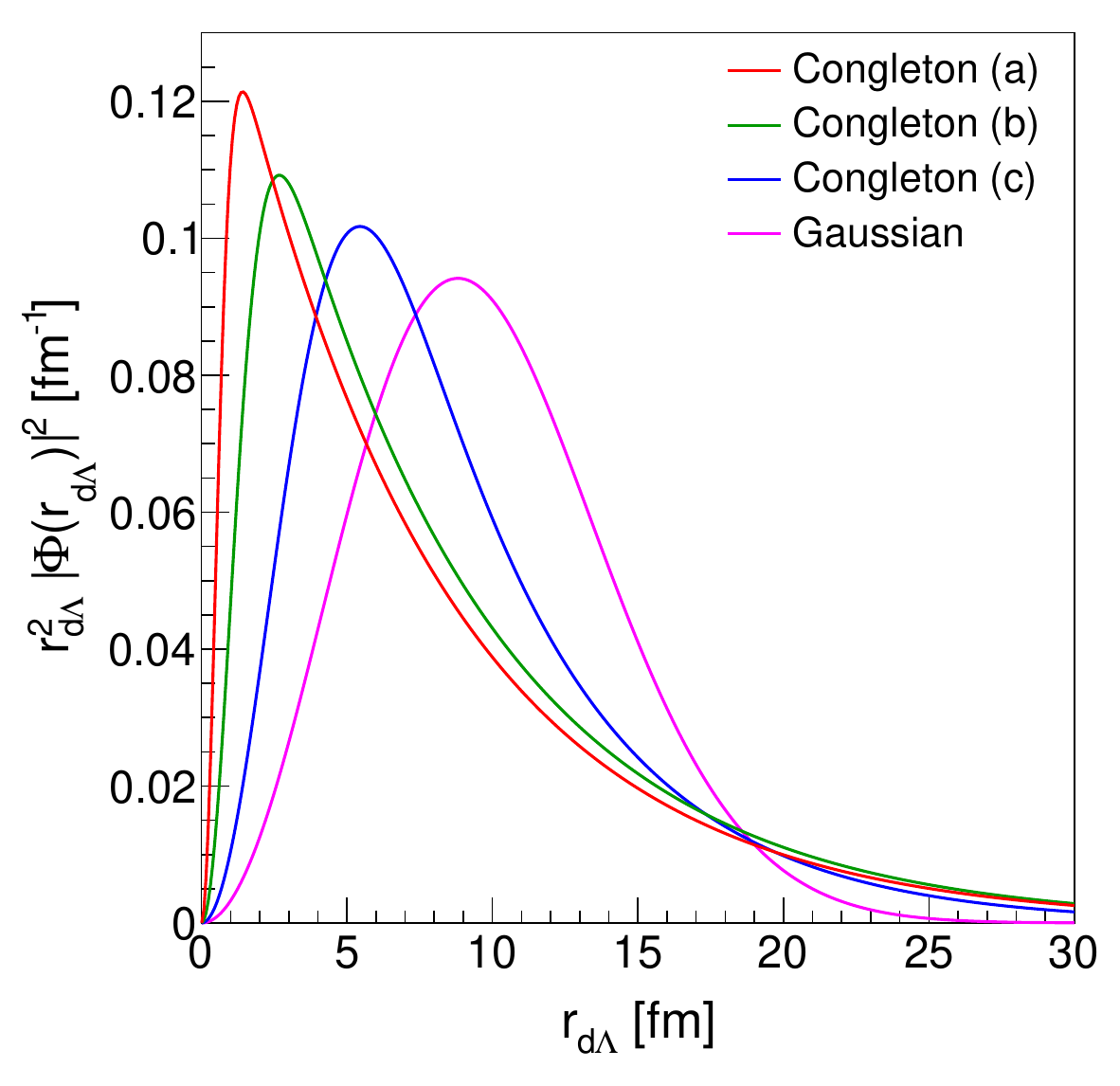} 
        \caption{The radial probability distributions, $r_{d\Lambda}^{2}  |\Phi(r_{d\Lambda})|^{2}$ of the ${}^{3}_{\Lambda}\rm{H}$ wave functions employed in this study. }
    \label{fig:rad_prob_congleton_wave_function}
\end{figure}

The radial probability distributions of the ${}^{3}_{\Lambda}\rm{H}$ wave functions shown in Fig.~\ref{fig:congleton_wave_function} are shown in Fig.~\ref{fig:rad_prob_congleton_wave_function}. 

\section{Thermal-FIST}
\label{app:thermal}

For the thermal model analysis, we employ the Thermal-FIST package~\cite{Vovchenko:2019pjl}, using settings analogous to those in Ref.~\cite{Reichert:2022mek}. The dependence of temperature and baryochemical potential on the collision energy, $T(\sqrt{s_{\rm NN}})$ and $\mu_{B}(\sqrt{s_{\rm NN}})$, is described through the chemical freeze-out curve from Ref.~\cite{Vovchenko:2015idt}, which is constrained by global data. A strangeness canonical ensemble with a strangeness correlation length of $3.55$ fm is adopted, as this choice provides the best description of the $\Lambda/p$ ratio at $\sqrt{s_{\rm NN}} = 3$ GeV~\cite{STAR:2024znc}. The strangeness suppression factor is fixed to $\gamma_{S} = 1$. The freeze-out radius $r(\sqrt{s_{\rm NN}})$ for $0$--$10\%$ central Au+Au collisions is determined following the procedure outlined in Ref.~\cite{Reichert:2022mek}. Specifically, we parameterize its collision energy dependence in the range $\sqrt{s_{\rm NN}} = 3$ GeV to $200$ TeV by fitting to STAR mid-rapidity data on charged pion multiplicities~\cite{STAR:2017sal}. The resulting function is given as:

\begin{align}\label{eq:pion}
\frac{dN_\pi^+}{dy} + \frac{dN_\pi^-}{dy} = a \times s_{\rm{NN}}^b \times \ln (s_{\rm{NN}}) - c
\end{align}

\noindent At each $\sqrt{s_{\rm{NN}}}$, we set $V(\sqrt{s_{\rm{NN}}})$ to a value such that the thermal model reproduces pion multiplicity from Eq.~\ref{eq:pion}. Table~\ref{tab:chemicalparameters} lists the chemical freeze-out parameters used in this study. 

\begin{table}[h]
    \centering
    \begin{tabular}{c c c c}
        \toprule
        $\sqrt{s_{\rm{NN}}}$ (GeV) & $T_{ch}$ (MeV) & $\mu_{B}$ (MeV) & R (fm) \\
        \midrule
        3     & 85.1  & 727.9  & 8.28  \\
        3.2   & 91.3  & 704.1  & 7.87  \\
        3.5   & 99.1  & 671.2  & 7.41  \\
        3.9   & 107.6 & 631.8  & 6.99  \\
        4.5   & 117.2 & 580.7  & 6.62  \\
        7.7   & 140.2 & 405.7  & 6.15  \\
        11.5  & 148.5 & 298.7  & 6.22  \\
        14.6  & 151.4 & 245.9  & 6.34  \\
        19.6  & 153.7 & 191.3  & 6.51  \\
        27    & 155.2 & 143.9  & 6.72  \\
        39	  & 156.1 & 102.7  & 6.96  \\
        54.5  & 156.5 &	75.0   & 7.18  \\
        200	  & 157.0 & 21.2   & 7.96  \\
        \bottomrule
    \end{tabular}
    \caption{Chemical freeze-out parameters used in this study.}
    \label{tab:chemicalparameters}
\end{table}

\section{Blast-Wave Predictions}

For the Blast-Wave analysis, the freeze-out parameters, namely the kinetic freeze-out temperature $T_{\rm kin}$, the mean radial velocity $\langle \beta \rangle$, and the radial flow profile parameter $n$, are taken from previously published fits to the spectra of light hadrons ($\pi^{\pm}$, $K(\bar{K})$, $p(\bar{p})$)~\cite{STAR:2008med, STAR:2017sal, STAR:2023uxk, E802:1999hit}. The values of the kinetic freeze-out parameters applied in this work are summarized in Table~\ref{tab:kineticparameters}.

\begin{table}[h]
    \centering
    \begin{tabular}{c c c c}
        \toprule
        $\sqrt{s_{\rm{NN}}}$ (GeV) & $T_{\rm kin}$ (MeV) & $\langle \beta \rangle$ ($c$) & $n$ \\
        \midrule
        3    & $63\pm10$   & $0.45\pm0.04$  & 1 \\
        4.8  & $127\pm15$  & $0.39\pm0.05$  & 1 \\
        7.7  & $117\pm11$  & $0.45\pm0.05$  & $0.5\pm0.3$ \\
        11.5 & $119\pm12$  & $0.46\pm0.05$  & $0.6\pm0.3$ \\
        19.6 & $114\pm12$  & $0.46\pm0.03$  & $0.9\pm0.2$ \\
        27   & $117\pm11$  & $0.48\pm0.03$  & $0.7\pm0.2$ \\      
        39   & $117\pm11$  & $0.48\pm0.04$  & $0.8\pm0.2$ \\
        62.4 & $100\pm11$  & $0.55\pm0.02$  & $0.7\pm0.4$ \\
        200  & $90\pm12$   & $0.59\pm0.05$  & $0.84\pm0.02$ \\
        \bottomrule
    \end{tabular}
    \caption{Kinetic freeze-out parameters used in the manuscript for blast-wave predictions.}
    \label{tab:kineticparameters}
\end{table}

\section{Coalescence calculations of triton spectra using nucleon source sizes obtained using different deuteron wave functions}
\label{sec:appendix_1}

\begin{figure}[h!] 
    \centering
    \includegraphics[width=.99\linewidth]{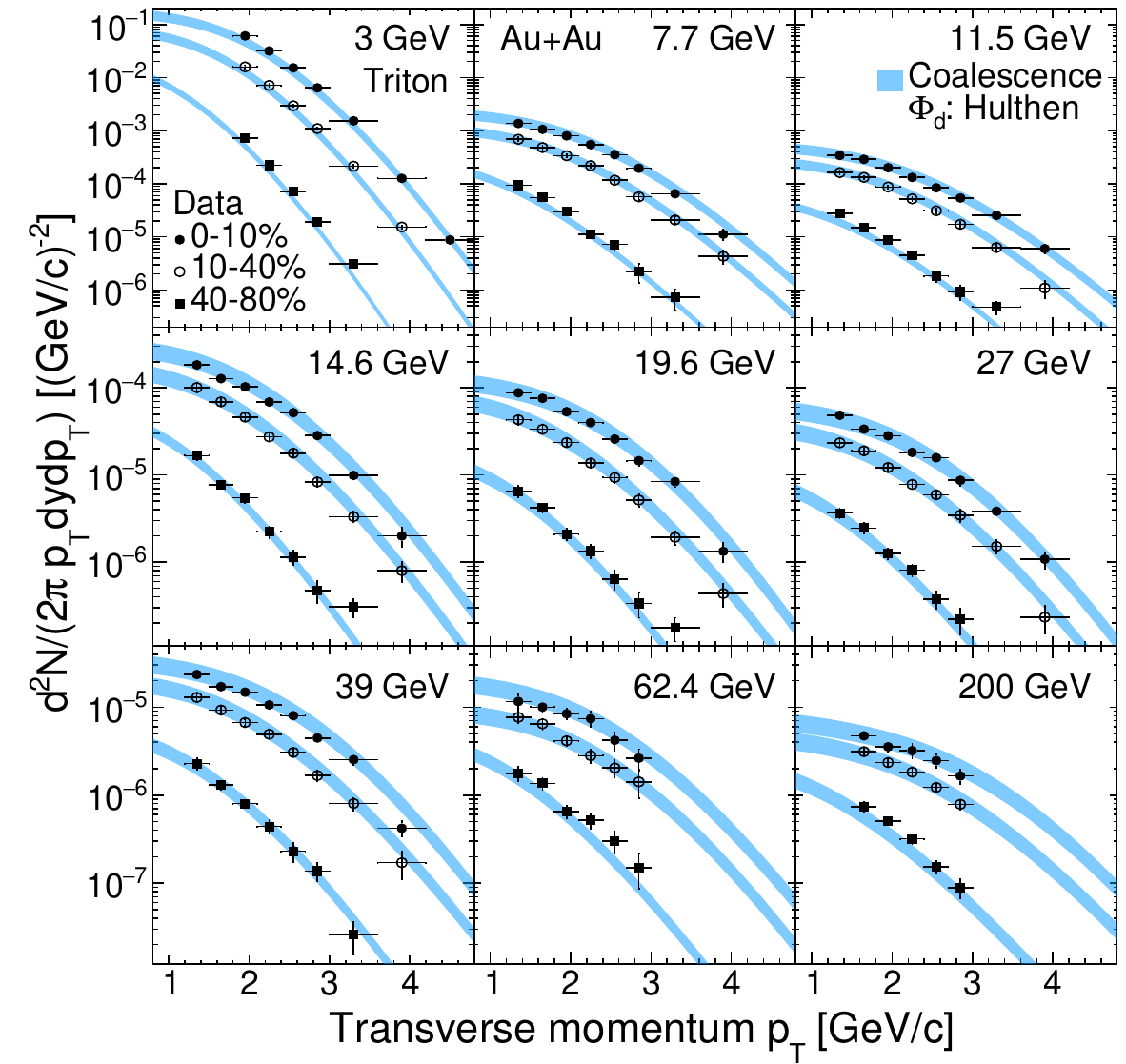} 
    \caption{Triton transverse momentum spectra at \snn=3, 7.7, 11.5, 14.6, 19.6, 27, 39, 62.4, and 200 GeV. The data~\cite{STAR:2022hbp} in three different centralities is compared with coalescence calculations. The Hulthen deuteron wave function is used in the source size calculation.}
    \label{fig:triton_spectra_coal_hulthen}
\end{figure}

\begin{figure}[h!] 
    \centering
    \includegraphics[width=.99\linewidth]{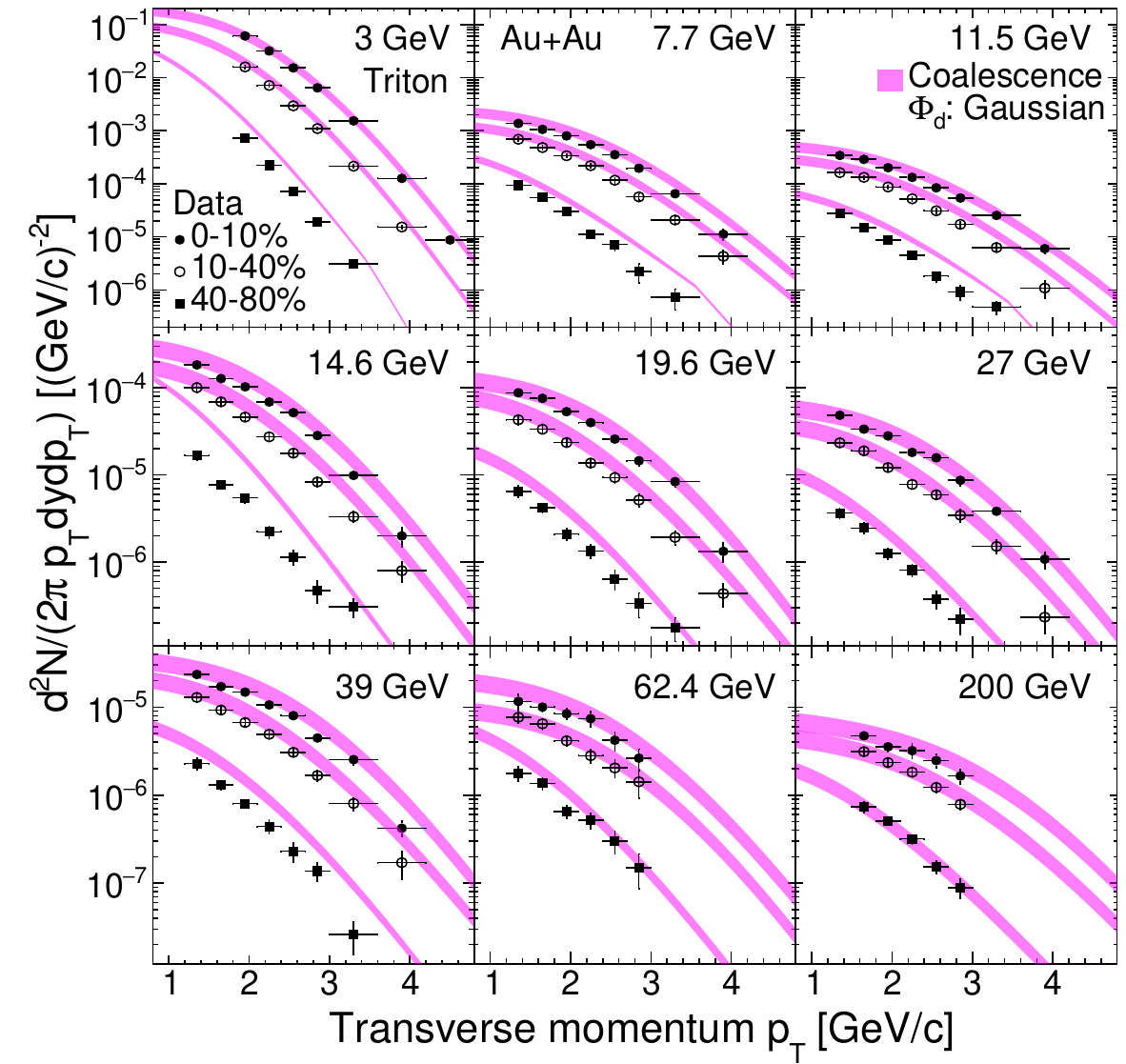} 
    \caption{Triton transverse momentum spectra at \snn=3, 7.7, 11.5, 14.6, 19.6, 27, 39, 62.4, and 200 GeV. The data~\cite{STAR:2022hbp} in three different centralities is compared with coalescence calculations. The Gaussian deuteron wave function is used in the source size calculation.}
    \label{fig:triton_spectra_coal_gaus}
\end{figure}

The coalescence calculations of triton spectra using nucleon source sizes obtained using the Hulthen wave function and the Gaussian wave function for the deuteron are shown in Fig.~\ref{fig:triton_spectra_coal_hulthen} and~\ref{fig:triton_spectra_coal_gaus} respectively.

\section{Coalescence calculations of $^{3}_{\Lambda}\rm{H}$ yields using wave functions with $\langle r_{d\Lambda} \rangle=9.1$~fm}
\label{sec:appendix_2}

The four wave functions employed in the main manuscript have $\langle r_{d\Lambda} \rangle \approx 10.8$~fm. The RMS deuteron–$\Lambda$ separation $\langle r_{d\Lambda} \rangle$ inferred from the ${}^{3}_{\Lambda}\mathrm{H}$ binding energy $0.163\pm0.036$~MeV~\cite{Liu:2024ygk, Eckert:2022dyz, Kasagi:2025mvh, ALICE:2022sco, STAR:2019wjm, Chaudhari1968, Juric:1973zq, Mayeur1966ADO, ammar1962, Prakash1961OnTB}, is $9.8^{+1.1}_{-0.7}~\text{fm}$. In this section, we explore wave functions with the lower limit $\langle r_{d\Lambda} \rangle=9.1$~fm. This is achieved by modifying the $b_{\Lambda}$ of the Gaussian wave function in Eq.~\ref{eq:gaussianhypertriton} from 7.2~fm to 6.1~fm, and the $\alpha_{\Lambda}$ of the Congleton wave functions (a), (b), and (c) in Eq.~\ref{eq:congleton} from 0.068~fm$^{-1}$, 0.068~fm$^{-1}$, and 0.090~fm$^{-1}$ to 0.082~fm$^{-1}$, 0.088~fm$^{-1}$, and 0.116~fm$^{-1}$ respectively while keeping $Q_{\Lambda}$ unchanged. 

\begin{figure}[h!] 
    \centering
    \includegraphics[width=.85\linewidth]{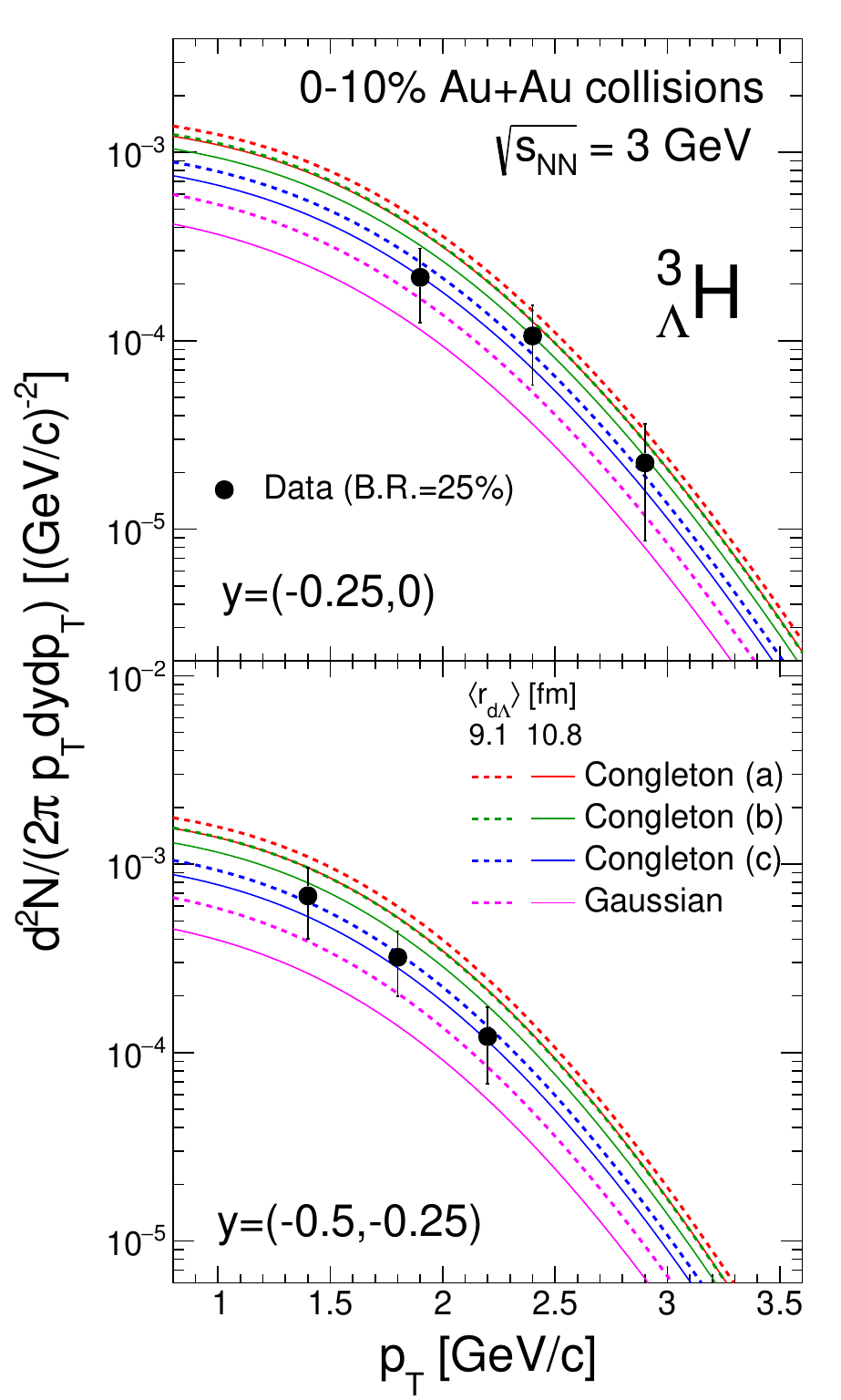} 
        \caption{${}^{3}_{\Lambda}\rm{H}$ transverse momentum spectra in $0-10\%$ \snn=3 GeV collisions. The data~\cite{STAR:2021orx}, in two rapidity regions $(-0.25,0)$ (upper panel) and $(-0.5,-0.25)$ (lower panel) are compared with coalescence calculations using different ${}^{3}_{\Lambda}\rm{H}$ wave functions are shown. The uncertainty on the calculations are not shown.}
    \label{fig:h3l_spectra_3gev_y4_update}
\end{figure}

Using this new set of $^{3}_{\Lambda}\rm{H}$ wave functions, we first calculate the $p_{T}$ spectra at $\sqrt{s_{\rm{NN}}}=3$ GeV. As shown in Fig.~\ref{fig:h3l_spectra_3gev_y4_update}, using more compact wave functions with $\langle r_{d\Lambda} \rangle=9.1$ fm lead to increased yields compared to those using wave functions with $\langle r_{d\Lambda} \rangle\approx 10.8$ fm. The increase is largely independent of $p_{T}$, and is largest for the Gaussian wave function, around $40\%$, and around $15-20\%$ for the Congleton wave functions. This is expected because the more compact wave functions enhance the probability at lower $r_{d\Lambda}$, thus increasing the overlap with the nucleon source at $\sqrt{s_{\rm{NN}}}=3$ GeV. 

\begin{figure}[h!] 
    \centering
    \includegraphics[width=.9\linewidth]{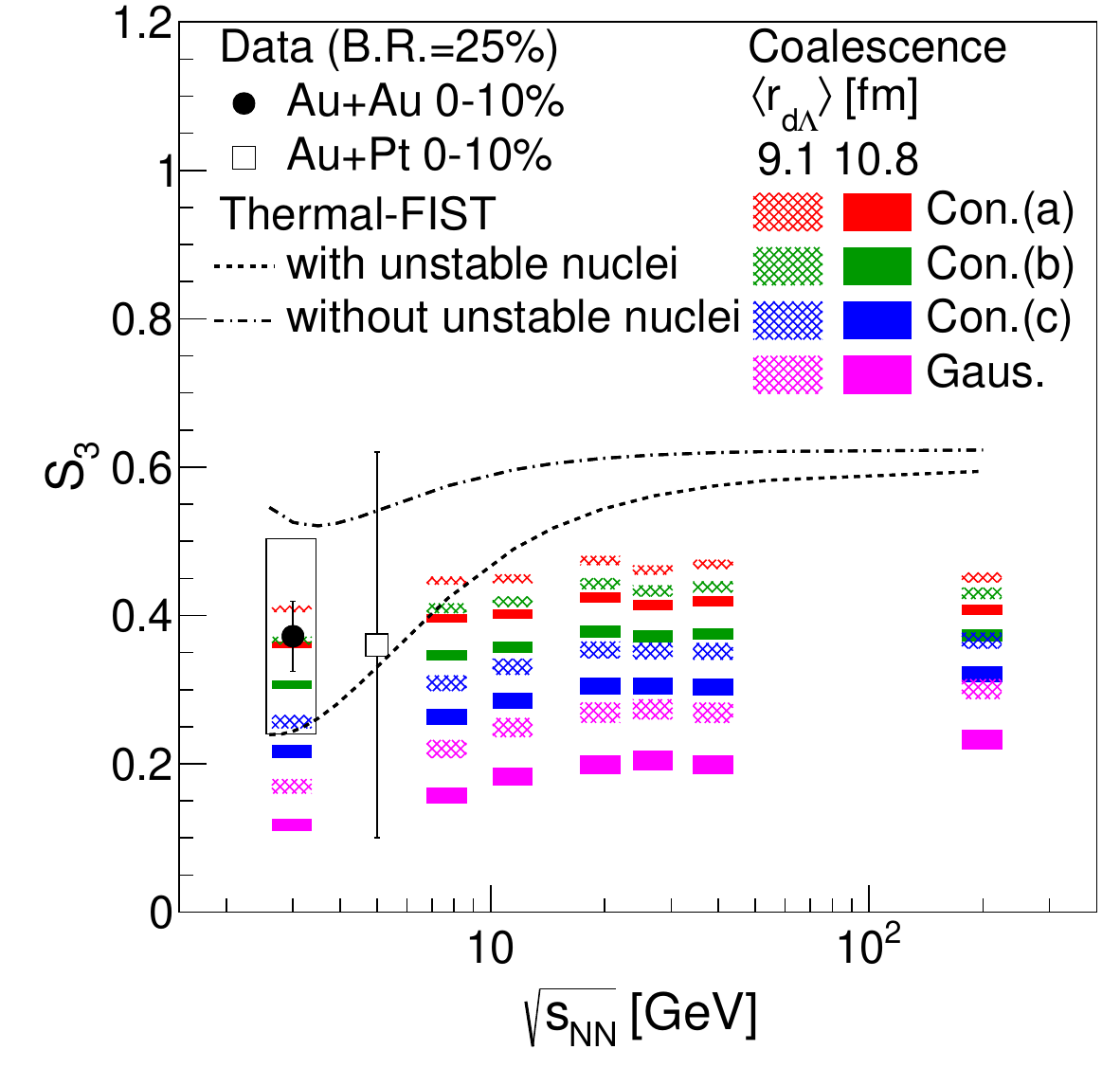} 
	\caption{$S_{3}$ in heavy-ion collisions as a function of $\sqrt{s_{\rm{NN}}}$. The data~\cite{STAR:2021orx, STAR:2023uxk, STAR:2024znc, E864:2002xhb} are compared with coalescence calculations using different ${}^{3}_{\Lambda}\rm{H}$ wave functions (colored bands) and Thermal-FIST calculations~\cite{Vovchenko:2019pjl} (dashed lines) including and not including unstable nuclei.}
    \label{fig:s3_vs_snn_update}
\end{figure}

\begin{figure}[h!] 
    \centering
    \includegraphics[width=.9\linewidth]{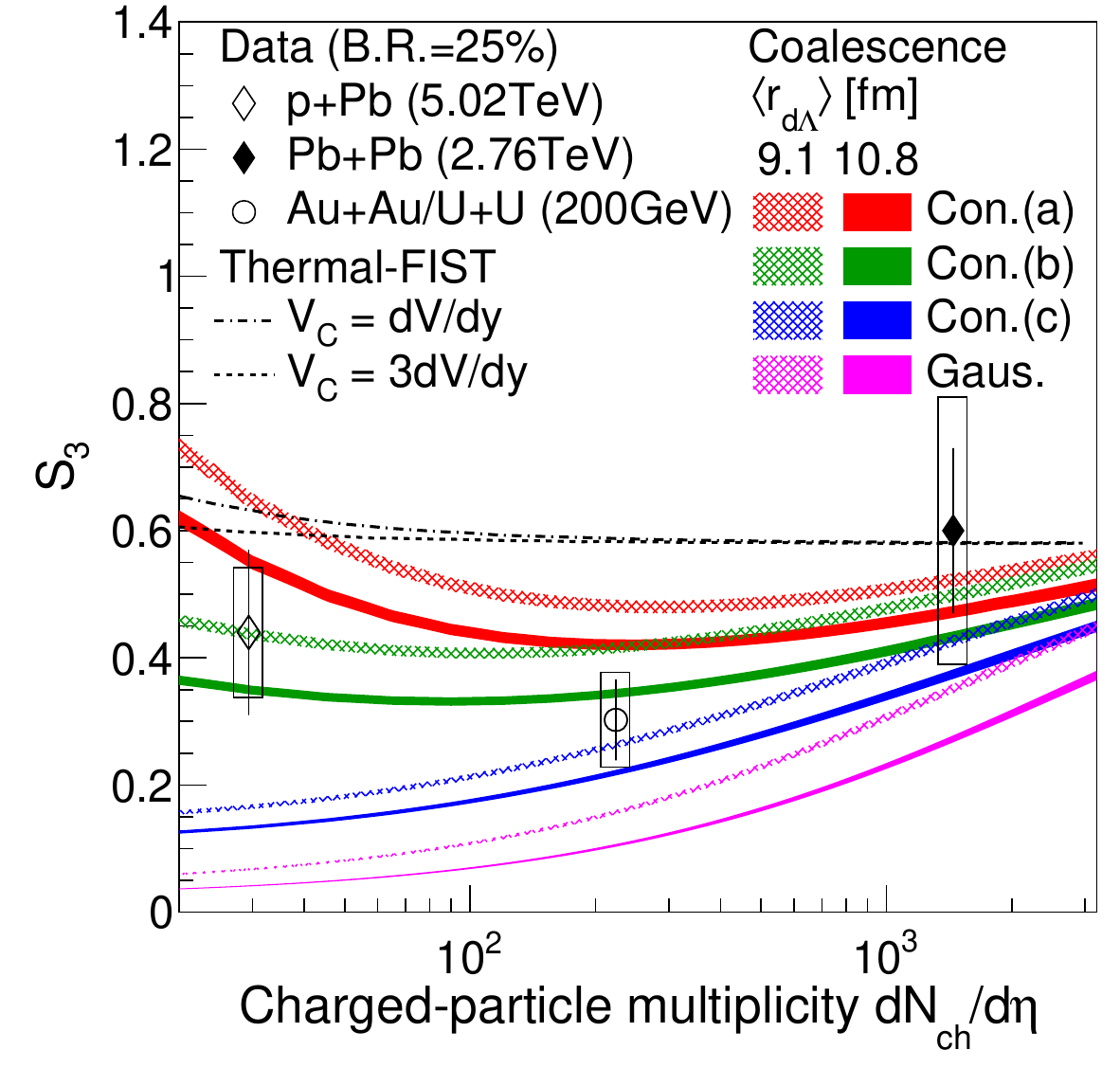} 
	\caption{$S_{3}$ in heavy-ion collisions as a function of charged-particle multiplicity $dN_{\rm{ch}}/d\eta$. The data~\cite{ALargeIonColliderExperiment:2021puh, STAR:2023fbc, ALICE:2015oer} are compared with coalescence calculations using different ${}^{3}_{\Lambda}\rm{H}$ wave functions (colored bands) and Thermal-FIST calculations~\cite{Vovchenko:2019pjl} (dashed lines).}
    \label{fig:s3_vs_nch_update}
\end{figure}

We then repeat the calculations using this new set of wave functions of $S_{3}$ as a function of collision energy and multiplicity, and the results are shown in Fig.~\ref{fig:s3_vs_snn_update} and~\ref{fig:s3_vs_nch_update} respectively. Similarly, we conclude that using more compact wave functions with $\langle r_{d\Lambda} \rangle=9.1$ fm lead to increased $S_{3}$ compared to those using wave functions with $\langle r_{d\Lambda} \rangle\approx 10.8$ fm. The relative increase is largest for low multiplicities and low energies; and becomes smaller at high multiplicities and higher energies. Nonetheless, the differences in $S_{3}$ between the results using different wave functions remain large in the low multiplicity range. 

\bibliographystyle{apsrev4-2} 
\bibliography{ref} 

\end{document}